\documentclass[acmtog,author]{acmart}
\usepackage{makecell}  
\usepackage{listings}
\usepackage[table]{xcolor}
\usepackage{cleveref}
\usepackage[normalem]{ulem} 
\usepackage{enumitem}
\usepackage{comment}
\usepackage{bm}
\usepackage{needspace}
\usepackage{wrapfig}
\usepackage{tablefootnote}
\usepackage{array} 
\usepackage{makecell}
\usepackage{pgfplots}
\usepackage{subcaption}
\pgfplotsset{compat=1.18}

\usepackage{caption}
\usepackage{tikz}

\usepackage[utf8]{inputenc}  
\DeclareUnicodeCharacter{0301}{\'{}} 

\usepackage{algorithm}
\usepackage{algpseudocode}
\usepackage{amsmath}

\usepackage{tcolorbox}
\tcbuselibrary{skins, breakable}
\usepackage{tabularx}

\newtcolorbox{graymathbox}[1][]{
  breakable,
  enhanced,
  colback=gray!2,       
  colframe=gray!2,      
  boxrule=0pt,          
  arc=4mm,              
  coltitle=black,       
  fonttitle=\bfseries,  
  title=#1,             
  top=0pt,
  bottom=1pt,
  left=4pt,
  right=4pt,
  before skip=0pt,
  after skip=2pt
}

\definecolor{lightgray}{RGB}{240,240,240}
\definecolor{darkgray}{RGB}{64,64,64}

\lstdefinestyle{pseudocode}{
    backgroundcolor=\color{lightgray},
    basicstyle=\ttfamily,
    commentstyle=\color{darkgray},
    keywordstyle=\bfseries,
    tabsize=2,
    breaklines=true,
    escapeinside={(*@}{@*)},
    frame=single,
}

\AtBeginDocument{%
  }
\setcopyright{acmlicensed}
\copyrightyear{2026}
\acmYear{2026}
\setcopyright{cc}
\setcctype{by}
\acmConference[SIGGRAPH Conference Papers '26]{Special Interest Group on Computer Graphics and Interactive Techniques Conference Conference Papers}{July 19--23, 2026}{Los Angeles, CA, USA}
\acmBooktitle{Special Interest Group on Computer Graphics and Interactive Techniques Conference Conference Papers (SIGGRAPH Conference Papers '26), July 19--23, 2026, Los Angeles, CA, USA}
\acmDOI{10.1145/3799902.3811095}
\acmISBN{979-8-4007-2554-8/2026/07}

\begin{document}
\citestyle{acmauthoryear}

\newcommand{\review}[1]{\textcolor{blue}{#1}}
\newcommand{\bing}[1]{\textcolor{orange}{[bing: #1]}}
\newcommand{\ravi}[1]{\textcolor{green}{[ravi: #1]}}
\newcommand{\marco}[1]{\textcolor{blue}{[marco: #1]}}
\newcommand{\mukund}[1]{\textcolor{teal}{[mukund: #1]}}
\newcommand{\lifan}[1]{\textcolor{red}{[Lifan: #1]}}
\newcommand{\bart}[1]{\textcolor{orange}{[bart: #1]}}
\newcommand{\todo}[1]{\marginpar{\Large {\color{red} $\spadesuit$}} {\color{red} [TODO: #1]}}

\newcommand{\edit}[1]{\textcolor{blue}{[edited: #1]}}

\newcommand{\bx}{\mathbf{x}}
\newcommand{\bom}{\boldsymbol{\omega}}
\newcommand{\bn}{\mathbf{n}}
\newcommand{\D}{\mathrm{d}}
\newcommand{\calH}{\mathcal{H}}
\renewcommand{\thefootnote}{\fnsymbol{footnote}}

\title{A Generalizable Light Transport 3D Embedding for Global Illumination}

\author{Bing Xu}
\email{binxu@nvidia.com}
\orcid{0009-0005-7359-8570}
\affiliation{%
  \institution{UC San Diego and NVIDIA}
  \country{USA}
}

\author{Mukund Varma T}
\email{tmukund@ucsd.edu}
\orcid{0000-0001-6480-3126}
\affiliation{%
  \institution{UC San Diego}
  \country{USA}
}

\author{Cheng Wang}
\email{chengwang@ucsd.edu}
\orcid{0009-0009-7630-3474}
\affiliation{%
  \institution{UC San Diego}
  \country{USA}
}

\author{Tzu-Mao Li}
\email{tzli@ucsd.edu}
\orcid{0000-0001-5443-470X}
\affiliation{%
  \institution{UC San Diego}
   \country{USA}
}

\author{Lifan Wu}
\email{lifanw@nvidia.com}
\orcid{0000-0002-5735-0998}
\affiliation{%
  \institution{NVIDIA}
   \country{USA}
}

\author{Bartlomiej Wronski}
\email{elirian@gmail.com}
\orcid{0009-0005-0806-2307}
\affiliation{%
  \institution{NVIDIA}
   \country{USA}
}

\author{Ravi Ramamoorthi}
\email{ravir@cs.ucsd.edu}
\orcid{0000-0003-3993-5789}
\affiliation{%
  \institution{UC San Diego}
   \country{USA}
}

\author{Marco Salvi}
\email{msalvi@nvidia.com}
\orcid{0009-0003-1366-2396}
\affiliation{%
  \institution{NVIDIA}
   \country{USA}
}

\begin{abstract}

Global illumination (GI) is essential for realism but remains computationally expensive. While per-scene neural methods lack generalization and screen-space approaches inherently suffer from view inconsistency, prior 3D neural rendering methods face a severe scalability barrier, restricting them to small, object-centric meshes.
To overcome these trade-offs, we introduce a generalizable light transport 3D embedding that predicts global illumination directly from 3D scene configurations without rasterized or path-traced illumination cues, per-scene retraining or screen-space limitations. We employ a point-based representation to decouple our embedding from the original scene topology, then utilize a linear-complexity transformer to encode long-range light transport.
This design scales to environments with millions of triangles, enabling the first generalizable GI learning on complex, high-fidelity indoor scenes, far beyond prior limits. 
To achieve this, we enforce a local query mechanism where rendering queries are processed independently under 3D supervision. This ensures constant complexity per pixel relative to scene size, yielding view-consistent and resolution-agnostic rendering without the memory bottlenecks typical of globally coupled attention. We further demonstrate versatility by re-targeting the encoder with limited fine-tuning, presenting preliminary results on spatial-directional radiance field prediction for glossy materials and validating transfer to downstream rendering tasks.

\end{abstract}


\begin{teaserfigure}
  \includegraphics[width=\textwidth]{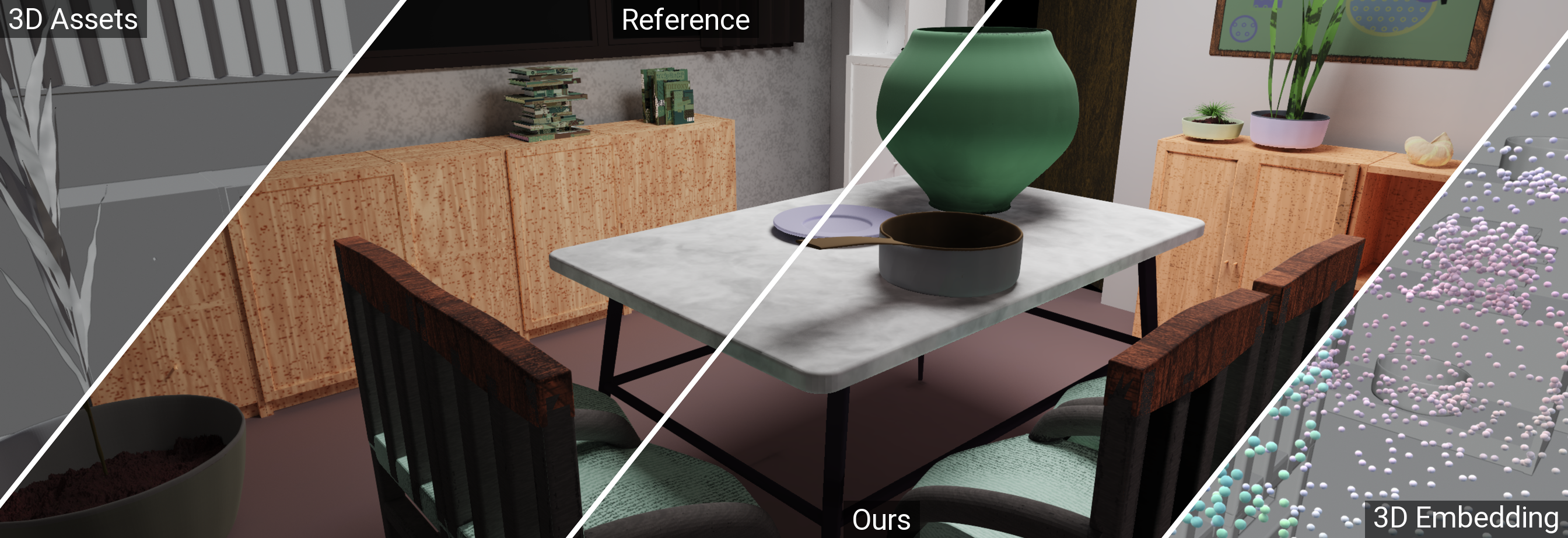}
   \caption{Our scalable transformer-based, generalizable light-transport model takes 3D scene assets (leftmost slice)—geometry, materials, and lighting as a point cloud—and encodes global light transport into a 3D embedding (rightmost slice) via latent codes anchored at scene points. The model is view- and resolution-independent. Center slices: reference vs. our predicted global illumination for an unseen scene at 2688×1152 resolution (cropped for space).}
  \label{fig:teaser}
\end{teaserfigure}

\maketitle

\section{Introduction}
A long-term goal of computer graphics has been to efficiently simulate complex light-matter interactions. 
While researchers have derived mathematical models for physical phenomena~\cite{christensen2016path}, recent years have witnessed a different paradigm: learning data-driven priors to approximate these processes~\cite{openai2024sora}. 
With the growing availability of 3D data~\cite{siddiqui2024assetgen}, we seek to answer a central question: can data priors alone be used to train a generalizable model that predicts 3D light transport directly from scene configurations, partially bypassing explicit simulation?

Monte Carlo rendering remains the workhorse of photorealism but is computationally intensive.
Recent neural global illumination methods~\cite{ren2013radiancereg, diolatzis2022active, rainer2022neuralpt, granskog2020compositional, zheng2023nelt} approximate light transport to enable efficient reuse of lighting computation, but they typically overfit to individual scenes, preventing generalization.
Conversely, cross-scene generalization efforts, including neural denoisers~\cite{afra2024openimagedenoise} or G-buffer–based predictors~\cite{nalbach2017deepshading}, often operate in 2D screen space. These methods suffer from view-inconsistency and fail to capture global illumination from off-screen geometry. 

Closer to our goal are methods that learn light transport directly in 3D. However, these approaches face a severe \textbf{scalability barrier}. Pioneering work by \citet{hermosilla2019deep} based on convolutions struggles to model long-range global interactions, and the recent transformer-based approaches like RenderFormer~\cite{zeng2025renderformer} rely on global attention with quadratic complexity. This computational bottleneck restricts them to simple, object-centric meshes, making them well-suited to object-centric setups but unsuitable for the dense geometry of real-world environments.

In this paper, we introduce a generalizable light transport 3D embedding designed to prioritize scalability and versatility. 
We achieve a scalable architecture by reducing computational complexity in three key ways:  1) \textit{Point-based primitives} decouple lighting resolution from the original scene topology and unify light sources with geometry into a single representation. 2) \textit{A linear-complexity transformer} replaces quadratic global attention to efficiently encode long-range light transport into the 3D embedding. 3) \textit{Independent local decoding.} We retrieve features only from a fixed local neighborhood ($k$-NN) of the embedding regardless of the scene size and process each query independently. This ensures no dependencies exist between separate rendering rays, enabling flexible training batch sizes and resolution-agnostic inference. Finally, we use direct 3D supervision in world space rather than 2D image space. This guarantees that our learned embedding is inherently view-consistent and robust to camera trajectory shifts.

In summary, we make the following contributions:
\begin{itemize}[leftmargin=*]
    \item A scalable 3D embedding using point clouds and linear-complexity transformers to overcome quadratic memory bottlenecks, enabling global transport modeling on millions of triangles (\S\ref{sec:method_primitives},\S\ref{sec:global_encoding}).
    \item Resolution-agnostic decoding: A local decoding mechanism with constant per-pixel complexity that, via 3D supervision, ensures view-consistent and generalizable GI for large-scale scenes (\S\ref{sec:local_query_decoding}).
    \item We curate and release a $\sim$14k complex indoor scenes dataset
    with diverse layouts, geometries, and textures, serving as a benchmark for learning light transport (\S\ref{sec:dataset}).
    \item Architectural versatility: A transferable encoder capable of supporting dedicated tasks, demonstrated by extending our primary diffuse GI model to predict spatial-directional radiance fields for glossy materials (\S\ref{sec:glossy}).

\end{itemize}

\definecolor{roadmapBlue}{RGB}{0, 114, 178}
\definecolor{roadmapOrange}{RGB}{230, 159, 0}

\section{Related work}
We focus our discussion on data-driven methods, particularly those leveraging neural networks to approximate light transport or operate directly on 3D data.




\paragraph{Per-scene neural rendering} Classical Precomputed Radiance Transfer (PRT)~\cite{sloan2002prt,ramamoorthi2009precomputation} and its neural extensions~\cite{rainer2022neuralpt, raghavan2023neural} enable efficient GI but rely on per-scene precomputation. Similarly, early neural regressions~\cite{ren2013radiancereg} and subsequent scene representations have focused on specific dynamic factors, such as moving objects~\cite{zheng2023nelt,zheng2024neural}, variable lighting or materials~\cite{gao2023neuralgi,diolatzis2022active}, and changing viewpoints~\cite{eslami2018neural,granskog2020compositional}. 
Hybrid neural renderers incorporate 3D scene reasoning via point-based light transport modules, trained per scene, while synthesizing final images in screen space~\cite{Sanzenbacher2020ARXIV}.
However, these methods remain fundamentally instance-specific. In contrast, our approach targets \emph{scene-level generalization}, learning a light transport embedding directly from raw 3D configurations without per-scene training.

\paragraph{Generalizable screen-space techniques}
Neural denoising and deep shading are two major approaches that aim for scene-level generalization. Denoising methods~\cite{bako2017kernel, vogels2018denoising, chaitanya2017interactive, xu2019adversarial, gharbi2019sample} operate as post-processing, while deep shading~\cite{nalbach2017deepshading, xin2022lightweight} infers shading from G-buffers. While efficient, their reliance on screen-space input leads to view inconsistencies and misses off-screen light transport effects.

\paragraph{Learning light transport in 3D}
A more robust alternative is to learn light transport directly in the 3D world space. \citet{hermosilla2019deep} pioneered this direction but relied on convolutions, which generally prioritize local features and struggle to model long-range global interactions. More recently, \citet{zeng2025renderformer} successfully applied global transformer architectures to mesh vertices, producing high-fidelity results for complex glossy and specular effects. While effective for single objects, their reliance on standard global attention introduces a quadratic computational bottleneck. This complexity constrains the approach to low-polygon scenes (e.g., $\approx$4k triangles), and is not the natural fit for dense, complex environments (see Figure~\ref{fig:scalability_renderformer}). Furthermore, our design adopts decoupled geometry-ray decoding, where each query independently attends to a local neighborhood of scene primitives, yielding view consistency and resolution flexibility. We provide a detailed structural comparison with \cite{zeng2025renderformer} in \S\ref{sec:model_ablation} and related statistics in the Supplementary \S3.
Looking ahead, a shared challenge for both methods is to further scale up with rendering parameters to capture the full spectrum of light transport effects.

\paragraph{Deep learning on irregular geometric data}
Processing irregular 3D data requires handling permutation invariance and varying density. Early MLP-based approaches~\cite{qi2017pointnet, qi2017pointnetplusplus} and subsequent graph~\cite{wang2019dynamic} or continuous convolution methods~\cite{thomas2019kpconv, li2019pointconv} proved effective for local geometry.
Attention‐driven architectures such as Point Transformer~\cite{zhao2021pointtransformer} further enhance the integration of global context. PointTransformerV3~\cite{wu2024pointtransformerv3} (PTV3) replaces per-layer KNN with serialization-based patch attention to scale to large point clouds; we adopt it as our encoder backbone. A brief background is provided in Supplementary \S1.2.
This choice prioritizes scalability and architectural simplicity, enabling efficient learning of complex relationships within 3D point clouds for global illumination.


\begin{figure}[!t]
    \centering
    \begin{tikzpicture}
\begin{axis}[
    width=1.0\linewidth, 
    height=3.6cm,
    font=\sffamily,
    xmin=0, xmax=46000,
    ymin=0, ymax=55,
    xtick={10000, 30000, 45000, 70000, 90000},
    xticklabels={10k, 30k, 45k, 70k, 90k},
    tick label style={font=\small\sffamily},
    legend style={font=\small\sffamily},
    ylabel near ticks,
    xlabel near ticks,
    grid=major,
    grid style={dashed, gray!30},
    legend pos=north west,
    scaled x ticks=false,
    axis line style={gray!30, line width=1.1pt},
    legend style={draw=none, fill=none, at={(0.2, 0.98)}, anchor=north,  cells={anchor=west}, font=\footnotesize\sffamily}
]
\addplot[
    color=roadmapOrange,
    mark=*,
    line width=1.1pt
]
coordinates {
    (5633, 12.07)
    (11266, 21.50)
    (16899, 30.97)
    (22532, 40.41)
};
\addlegendentry{RenderFormer}

\addplot[
    color=roadmapBlue,
    mark=square*,
    mark size=1.4pt,
    line width=1.1pt,
]
coordinates {
    (5633, 3.34)
    (11266, 4.04)
    (16899, 4.69)
    (22532, 5.36)
    (28165, 5.99)
    (33798, 6.67)
    (39431, 7.32)
    (45064, 8.00)
};
\addlegendentry{Ours}

\node[
        label={[text=roadmapOrange, align=center, font=\bfseries\footnotesize\sffamily, yshift=2pt]0:{OOM (Crash)}},
        mark size=5pt,
        color=roadmapOrange
    ] at (axis cs:28165, 50) {\pgfuseplotmark{x}};

\draw[roadmapOrange, dashed, thick] (axis cs:22532, 40.41) -- (axis cs:28165, 50);
\end{axis}
\end{tikzpicture}
\caption{\textsc{Memory scalability}. We benchmark \textit{Peak Training VRAM (GB)} (y) with increasing \textit{Number of primitives} (x) on a single scene using a tiny $\bf{64^2}$ \textbf{resolution}. Even with this reduced query count, RenderFormer rapidly exhausts a 48GB GPU. See more discussions in \S\ref{sec:scalability} and supplementary.}

\label{fig:scalability_renderformer}
\end{figure}

\section{Light transport 3D embedding}
\label{sec:embedding}
We introduce a generalizable framework for predicting global illumination directly from 3D configurations. Inspired by the parallel between the light transport operator and attention (\S\ref{sec:motivation}), we detail our pipeline's three key stages, as illustrated in Figure \ref{fig:pipeline}: Scene \emph{Discretization} (\S\ref{sec:method_primitives}), \emph{Global Light Transport Encoding} (\S\ref{sec:global_encoding}) where we encode these points into a high-dimensional neural representation ($\mathbf{F}_{i}$) that captures the light transport function of the given scene, and \emph{Local Query Decoding} (\S\ref{sec:local_query_decoding}). 

\subsection{Motivation: An Analogy Bridging Light Transport and Attention} \label{sec:motivation}  
\begin{figure}[!htbp]
    \centering
    \includegraphics[width=\linewidth]{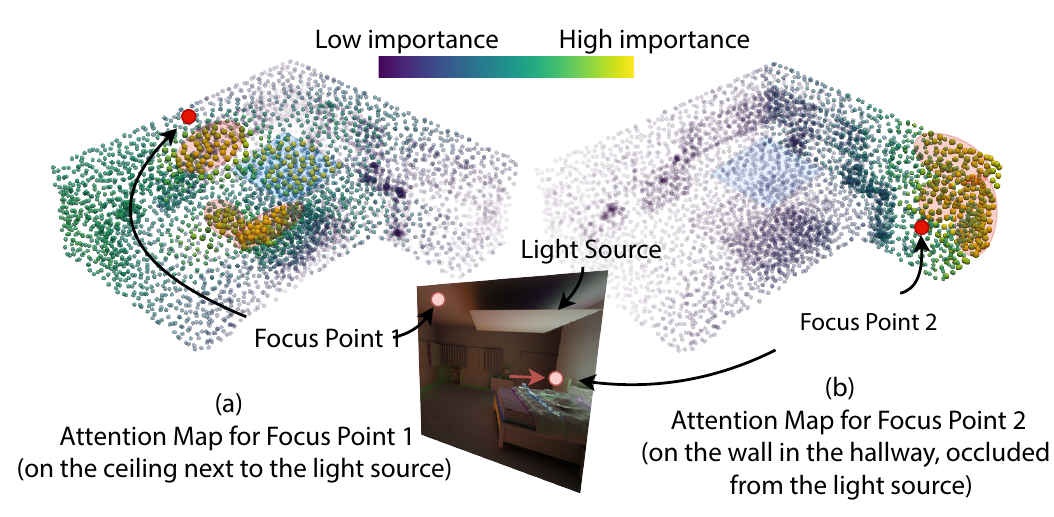}
    \caption{\textsc{Attention visualization for a bedroom with a hallway.} We show attention heatmaps from our light transport encoder for two focus points. Points are shaded by importance, with the top 200 scores enlarged. The blue shaded square denotes the light source. In (a), we see four groups of scene points having the largest impact on the focus point (highlighted in blue square and orange clusters), which likely corresponds to 4 bounces of ray tracing. In (b), the focus point is on the wall of the  hallway where the light source is occluded. This serves as a visual hint for our analogy between the light transport operator and the attention mechanism in \S\ref{sec:motivation}. } 
    \label{fig:attention_visualization}
\end{figure}

Light transport is often expanded as a Neumann series~\cite{arvo1994framework}, where global illumination is resolved by recursively propagating energy between surface points. This mirrors the stacked self-attention operation in transformers~\cite{vaswani2017attention}. Just as the transport operator $T$ (\citet{soler2022theoretical}) recursively simulates bounces, passing features through successive transformer layers allows the network to capture increasingly complex, multi-bounce interactions.
We offer a visual intuition for this parallel in Figure~\ref{fig:attention_visualization}, illustrating that attention weights naturally highlight scene regions contributing to a point's illumination.

Guided by this analogy, we design our Encoder to act as the global solver: by stacking deep attention layers, it ``bakes'' the multi-bounce simulation results into per-point embeddings. This allows our Decoder to simplify to a \textbf{local retrieval task}, avoiding the expensive global attention used in prior work~\cite{zeng2025renderformer} and enabling resolution-independent inference.


\subsection{Points as the Intermediate Scene Representation}
\label{sec:method_primitives}

Unlike prior generalizable methods that rely on fixed mesh topology (e.g., vertices) or view-dependent 2D projections~\cite{nalbach2017deepshading}, we adopt point clouds as our intermediate representation (IR), deriving per-point attributes from the 3D scene.
This design offers key advantages: 1) scalability: point density can adjust based on importance; 2) generality: decoupled from original geometry, suitable for other sources like scanned data; and 3) simplicity: easier for neural networks to handle than meshes, with no connectivity.

\paragraph{Light sources} While most prior neural GI works focus on distant illumination (e.g., environment maps), this approximation fails to capture near-field effects. We instead use local area lights to produce rich, spatially varying illumination essential for indoor rendering. 
By representing light sources as point clouds, we unify the treatment of emissive and non-emissive geometries. This makes our method agnostic to the lighting type, effortlessly supporting area lights, point lights, or emissive strips without architectural changes.

\paragraph{Scene points} 
As a pre-processing step (Figure~\ref{fig:pipeline}, left), we convert the scene into an intermediate representation by sampling object geometry into $M$ ($\approx 20k$) \textit{Scene points}. Each point $i$ is defined as a tuple $(\mathbf{p}_{i}, \mathbf{n}_{i}, \mathbf{c}_{i}, \mathbf{e}_{i})$, representing position, normal, albedo, and emissivity, respectively. The full scene is thus denoted as $\{(\mathbf{p}_{i}, \mathbf{n}_{i}, \mathbf{c}_{i}, \mathbf{e}_{i})\}^{M}$. This approach aligns with prior light transport approximations that also leverage point cloud inputs~\cite{havsan2006direct, hermosilla2019deep}. Note that the mesh vertices used in ~\citet{zeng2025renderformer} can be viewed as a specific realization of this representation. Further analysis of point sampling is provided in \S\ref{sec:model_ablation}.

\paragraph{Query points}
At render time, \textit{Query points} correspond to ray-intersections for shading, while during training, they serve as locations for loss evaluation. Each point is defined by features $\{(\mathbf{p}_{j}, \mathbf{n}_{j}, \mathbf{c}_{j})\}$. Crucially, rather than using view-dependent path tracing samples for supervision like many prior works (e.g. \cite{zeng2025renderformer}, which is limited by fixed resolutions and views), we uniformly sample $N$ ($\approx 2$ million) points from the scene surfaces. This approach decouples training from specific camera angles, enabling view and resolution independence. Further analysis is in \S\ref{sec:model_ablation}. \textit{Note we differentiate \emph{Scene points} from \emph{Query points} throughout the paper}.

\subsection{Global Light Transport Encoding}
\label{sec:global_encoding}
\begin{figure*}
    \centering
    \includegraphics[width=0.92\textwidth]{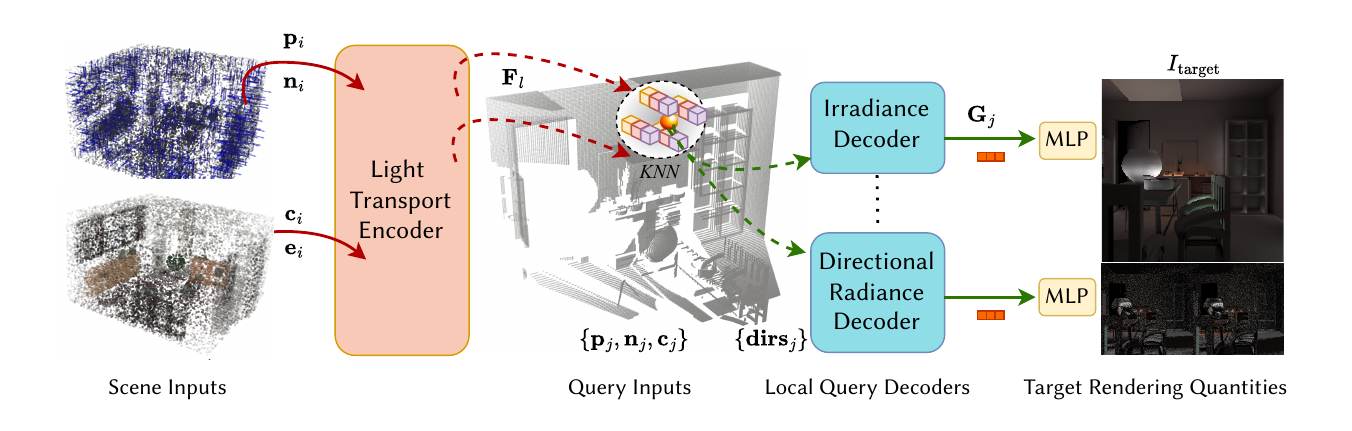}
    \caption{\textsc{Our pipeline}. Our method first converts the original 3D assets of each scene into a point cloud with associated features (on the left). The scenes are then encoded into a light transport embedding, consisting of latent codes (depicted as small blocks within the dashed circle) anchored at the scene point positions. At render time (center), each query point gathers local latent codes via k-nearest neighbors and feeds them together with the query point attributes into a target-specific query decoder, of which the core is a cross-attention. Note that direct emission can be added on top of the predicted output, so it needs not be a query input.}
    \label{fig:pipeline}
\end{figure*}

Our goal is to approximate global illumination effects directly from 3D scene inputs---geometries, materials, light sources---without relying on any rasterized or path-traced illumination cues.  
GI requires simulating the full complexity of light transport, including multiple bounces against various surfaces and interactions with various material properties (BSDFs), before reaching the camera.
Our observation from \S\ref{sec:motivation} motivates us to leverage a transformer-based encoder (dubbed \textit{Light Transport Encoder} in Figure~\ref{fig:pipeline}) to derive per-point neural primitives that implicitly model the light transport operator. 
As illustrated in Figure~\ref{fig:pipeline}, we encode these \emph{Scene points} into a high-dimensional neural representation ($\mathbf{F}_{i}$) that captures the light transport function of the given scene.

Given that a large number of points ($M \approx$ 20K) is required to represent an entire scene with sufficient density, na\"ively borrowing the original transformer architecture leads to prohibitive memory requirements. This is because self-attention scales quadratically, necessitating the storage of massive $M \times M$ attention maps. 
Rather, we leverage the state-of-the-art point cloud encoder PointTransformerV3~\cite{wu2024pointtransformerv3}, to derive a latent representation for each point. 

Our experiments indicate that na\"ively feeding all input features into the transformer yields suboptimal results. 
We attribute this to the attention operation becoming increasingly sparse with a large number of input tokens, which complicates the modeling of complex interactions between scene points.
Yet, aggressively reducing the number of input points can lead to information loss, as a large indoor scene cannot be adequately represented by a limited set of points.

\begin{figure*}
    \centering
    \includegraphics[width=0.99\textwidth]{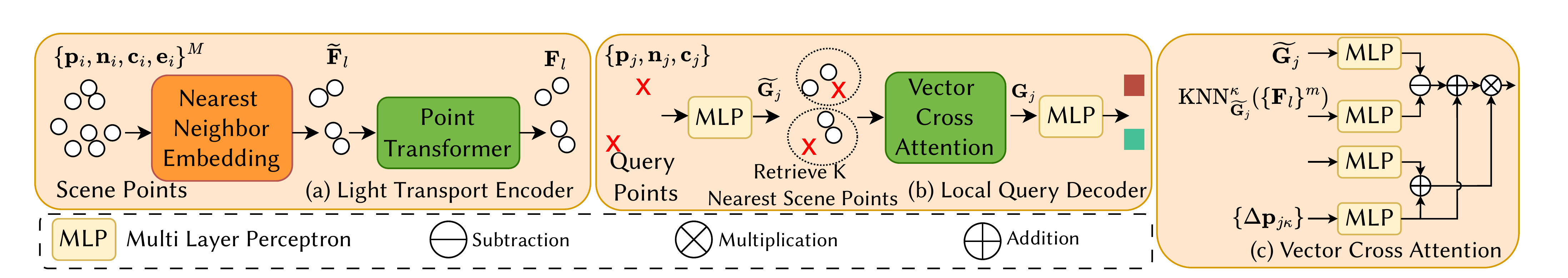}
      
    \caption{Network architecture for Light Transport Encoder and Local Query Irradiance Decoder.}
    \label{fig:network_arch}
\end{figure*}

Therefore, we derive an embedding from a sparser set of points, while also aggregating features from their immediate neighbors. The primary purpose of this Nearest Neighbor Embedding (NNE) is to pre-aggregate local features and increase embedding capacity before the transformer stage, rather than merely reducing the point count.
This mitigates the attention operation's ``learning overhead'' associated with large token counts while still preserving information from the full point set. We found that this hierarchical aggregation outperforms a flat PointTransformerV3 baseline by providing a more enriched latent representation. Formally, we write the Nearest Neighbor Embedding (see Figure~\ref{fig:network_arch}(a)) operation as follows:
%
\begin{align}
    \{\widetilde{\mathbf{X}}_{l}\}^{m} = \operatorname{FPS}(\{\mathbf{X}_{i}\}^{M}),\textit{ where }
    \mathbf{X}_{i} = \mathcal{F}((\mathbf{p}_{i}, \mathbf{n}_{i}, \mathbf{c}_{i}, \mathbf{e}_{i})) \\
    \widehat{\mathbf{X}}_{l, k} = \operatorname{concat}\left(\mathbf{X}_{k} - \widetilde{\mathbf{X}}_{l}, \widetilde{\mathbf{X}}_{l}\right) \quad \forall \mathbf{X}_{k} \in \operatorname{KNN}_{\widetilde{\mathbf{X}}_{l}}^{k}(\{\mathbf{X}_{i}\}^{M})\\
    \widetilde{\mathbf{F}}_{l} = \operatorname{max}_{k}\mathcal{G}(\widehat{\mathbf{X}}_{l, k})
\end{align}
%
\noindent The embedding process proceeds in three stages. Here in the first equation, $\mathcal{F}$ is a projection function that maps the raw input attributes of each point $i$ to a latent feature $\mathbf{X}_{i}$. We then utilize Farthest Point Sampling ($\operatorname{FPS}$) to subsample the original scene points $\{\mathbf{X}_{i}\}^{M}$ to $m$ points $\{\widetilde{\mathbf{X}}_{l}\}^{m}$. In our implementation, FPS is applied twice as a feature pre-aggregation step, halving the point count with each iteration (resulting in $m = M / 4)$, though this reduction factor remains flexible and can be adjusted as needed. After this step, we obtain subsampled point latents $\widetilde{\mathbf{X}}_{l}$. Next, we perform local feature aggregation. 
$\operatorname{KNN}_{\widetilde{\mathbf{X}}_{l}}^{k}$ identifies the $k$ nearest neighbors for each subsampled point $\mathbf{X}_{l}$ from the original scene, and the difference vectors between the KNN center and its neighbors are computed. 
These are then concatenated with each center $\widetilde{\mathbf{X}}_{l}$ using $\operatorname{concat}$, processed by a second projection $\mathcal{G}$, and then max-pooled along the $k$ neighbors to yield the final aggregated embedding $\widetilde{\mathbf{F}}_{l}$. 

As mentioned, we model interactions between these scene points using Point Transformer V3 (PTV3, \cite{wu2024pointtransformerv3}) as our backbone:
\begin{align}
    \{\mathbf{F}_{l}\}^{m} = \operatorname{PTV3}(\{\widetilde{\mathbf{F}}_{l}\}^{m}),
\end{align}

\noindent where $\operatorname{PTV3}$ operates on the derived scene embedding across all points $\{\widetilde{\mathbf{F}}_{l}\}^{m}$ to derive the final per-point latent $\{\mathbf{F}_{l}\}^{m}$ (dubbed \textit{Light Transport Embedding}). While standard point encoders typically rely on expensive $k$-Nearest Neighbor (KNN) searches per layer, PTV3 introduces \textit{point cloud serialization}. It orders points along a space-filling curve and computes attention within non-overlapping patches of consecutive points (using a patch size of 1024). This significantly reduces memory and latency; however, because attention is local within each patch, the spatial coverage per patch limits how much 3D context a single layer can directly reach. Our NNE pre-aggregation is specifically designed to complement this. With FPS applied twice ($m{=}M/4$) and $k{=}32$ neighbors per stage, each NNE-stage-1 point summarizes ${\sim}k$ raw points and each stage-2 point summarizes ${\sim}k$ stage-1 points, so each NNE point fed into the transformer accumulates context from up to ${\sim}k^2{=}1024$ raw scene points. A PTV3 patch of 1024 such ``enriched'' points therefore touches a massive 3D region (nearly $1/5$ of the entire scene at $M{=}20$k) that no patch of 1024 raw points could reach in a single attention operation. See Supplementary \S1.2 for an extended background.
The overall operation in this stage is denoted as the Light Transport Encoder in Figure~\ref{fig:network_arch}(a).

\subsection{Local Query Decoding}
\label{sec:local_query_decoding}

Our framework adopts a modular design, allowing the light-transport embedding to be repurposed for various rendering tasks by simply swapping task-specific decoders. We show that, after training on a large-scale indoor dataset of diffuse scenes for irradiance prediction (\S\ref{sec:diffuse}), the model can be fine-tuned to handle glossy reflections with limited additional training on a small glossy dataset, leveraging the pretrained weights (\S\ref{sec:glossy}).

Given the Light Transport Embeddings $\{\mathbf{F}_{l}\}^{m}$ of a scene, we wish to estimate global illumination quantities for arbitrary 3D points. For view-independent training, we sample the scene; for rendering, we obtain the ray intersection points, which we define as \textit{query points} (see \S\ref{sec:method_primitives} and Figure  \ref{fig:network_arch}). To do so, we propose a \emph{Local Query Decoder} that aggregates information from neighboring scene embeddings to synthesize the final desired output (e.g., irradiance).

Consistent with the previous section, we project each query point $j$ characterized by its position, normal, and material properties, i.e. $(\mathbf{p}_{j}, \mathbf{n}_{j}, \mathbf{c}_{j})$, into a latent feature $\widetilde{\mathbf{G}}_{j}$ using an MLP $\mathcal{H}$ (as in the equation below). 
Simultaneously, we retrieve the $\kappa$ nearest neighbors from the scene embeddings $\{\mathbf{F}_{l}\}^{m}$, denoted as $\mathbf{KV}$. Since neighbor importance varies based on distance and occlusion, simple averaging is insufficient; we therefore employ a learned aggregation strategy.
We verify this experimentally in \S\ref{sec:model_ablation}. To encode the spatial relationship, we compute the relative distance vector $\Delta \mathbf{p}_{j\kappa}$ between the query and each neighbor, mapping it to a high-dimensional positional encoding $\mathbf{P}_{j\kappa}$ via $\gamma$:

\begin{align}
   \widetilde{\mathbf{G}}_{j} = \mathcal{H}(\mathbf{p}_{j}, \mathbf{n}_{j}, \mathbf{c}_{j}) \\
   \mathbf{KV} = \operatorname{KNN}_{\widetilde{\mathbf{G}}_{j}}^{\kappa}(\{\mathbf{F}_{l}\}^{m})\\
    \mathbf{P}_{j\kappa} = \mathcal{\gamma} (\Delta \mathbf{p}_{j\kappa}), \Delta \mathbf{p}_{j\kappa} = \mathbf{p}_{j} - \mathbf{p}_{\kappa}.
\end{align}
\noindent As shown in Figure~\ref{fig:network_arch}(b), this explicitly provides both the local query context ($\widetilde{\mathbf{G}}_{j}$) and the relative geometry ($\mathbf{P}_{j\kappa}$) needed to weigh the contributions of neighboring scene points.

\paragraph{Vector Cross-Attention.} We leverage a transformer to iteratively aggregate features from neighborhood \textit{scene points} onto the \textit{query point}.
Crucially, rather than applying the standard dot-product attention between the query and scene points (as query and key, values respectively), we use a vector cross-attention operation. 
We replace the dot product attention with subtraction as the relation function.
Unlike the dot product, which collapses feature dimensions into a single scalar weight, subtraction-based attention computes distinct scores for each channel of the value matrix~\cite{zhao2021pointtransformer, fan2022cadtransformer}. This preserves channel-wise granularity and significantly increases the diversity of feature interactions when aggregating neighborhood information.
Additionally, we augment the attention and value matrices with the relative geometry $\mathbf{P}_{j\kappa}$ derived earlier.

\begin{flushleft}
\begin{align}
    \mathbf{G}_{j}^{q} = \operatorname{W}_{q} (\widetilde{\mathbf{G}}_{j}), \mathbf{G}_{j}^{k} = \operatorname{W}_{k} (\mathbf{KV}), \mathbf{G}^{v} = \operatorname{W}_{v} (\mathbf{KV})\\
    \mathbf{A} = \mathbf{G}_{j}^{q} - \mathbf{G}_{j}^{k} + \mathbf{P}_{j\kappa}\\ \mathbf{G}_{j} = \operatorname{sum}_{\kappa} (A (\mathbf{G}^{v} + \mathbf{P}_{j\kappa})) 
\end{align}
\end{flushleft}

\noindent Formally, as shown in the equations, the learned linear projections $\operatorname{W}_{q}$ ,$\operatorname{W}_{k}$ and $\operatorname{W}_{v}$ 
project the query ($\widetilde{\mathbf{G}}_{j}$), key, and value inputs ($\mathbf{KV}$) into their respective latent spaces.
\noindent The final embedding for each query point, denoted by $\mathbf{G}_{j}$, is obtained by aggregating the value features using the computed attention values $\mathbf{A}$ through summing up along the $\kappa$ neighboring points (see Figure~\ref{fig:network_arch}(b) and (c)).

The final output $\mathbf{I}_{\text{out}}$ is derived as follows:
\begin{align}
     \mathbf{I}_{\text{out}} = \mathcal{W}_{\text{out}}(\{\mathbf{G}_{j}^{N}\}),
\end{align}
\noindent where $\mathcal{W}_{\text{out}}$ represents the final output projection to convert the set of query points $\{\mathbf{G}_{j}^{N}\}$ to the required rendering quantities. These projections are implemented as multi-layer perceptrons.


\section{Implementation}
\begin{figure}[!t]
    \centering
    \includegraphics[width=0.9\linewidth]{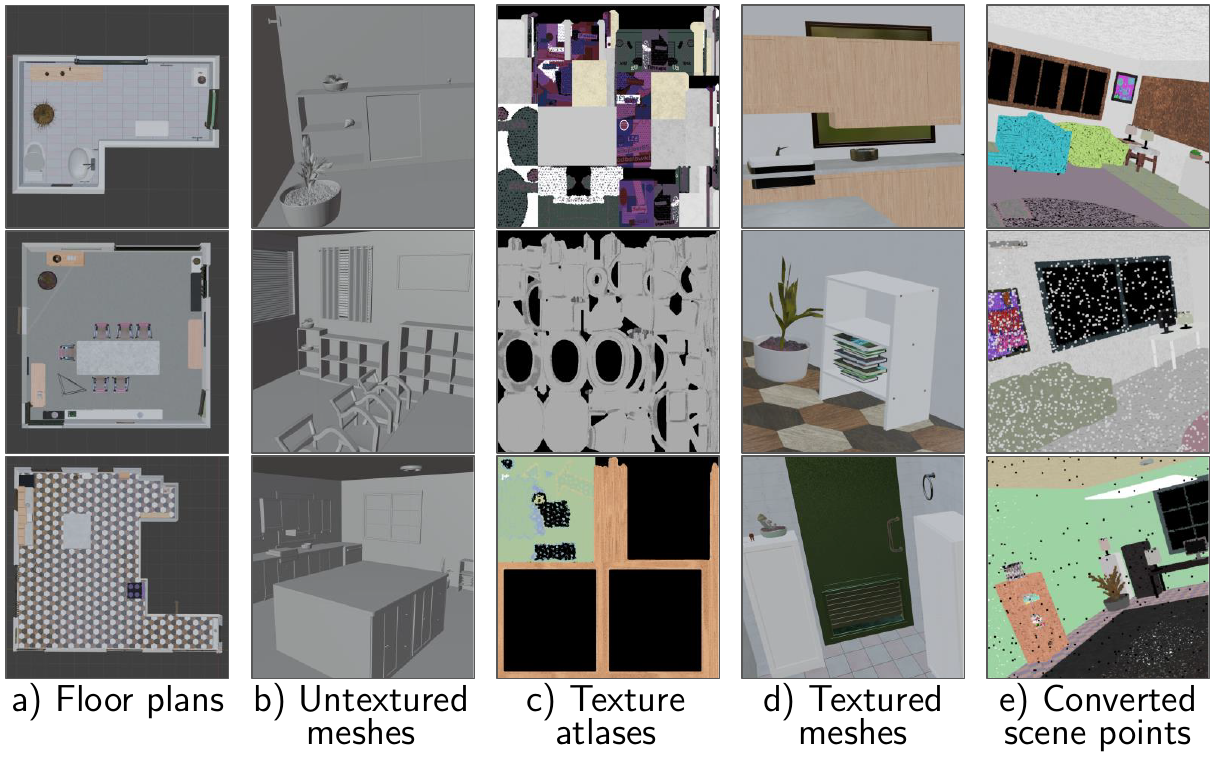}
    \caption{\textsc{Dataset overview.} Examples of: a) generated floor plans, b) untextured meshes for a floor plan, c) texture atlases, d) textured mesh objects, e) textured meshes converted to scene points directly fed into our pipeline.}
    \label{fig:dataset_overview}
\end{figure}
\paragraph{Training dataset}
\label{sec:dataset}
We curated a large-scale indoor global illumination dataset comprising 13,900 procedural scenes~\cite{raistrick2024infinigen} spanning diverse floorplans and asset collections (\Cref{fig:dataset_overview}). To support future research, we will publicly release all data, code, and trained weights.
We generated high-quality ground-truth irradiance by path-tracing approximately 2 million query points per scene (1,024 rays/point). Additionally, we curated a focused subset featuring glossy materials modeled with the Trowbridge–Reitz (GGX) distribution~\cite{walter2007microfacet} to validate our versatility experiments. Detailed data generation protocols, including geometry refinement and lighting heuristics, are provided in the Supplementary \S1.



\paragraph{Training scheme: 3D supervision}
\label{sec:training_scheme}
To avoid dependence on specific camera views, we train directly on query points uniformly sampled across the entire scene—roughly 2 million points per scene. 
We train on 90\% of the 12k scenes and hold out the rest to test the generalizability. Our model is trained on four A10 GPUs for approximately seven days, using learning rate warm-up followed by cosine decay. Each batch contains 8,192 query points, with a per-GPU batch size of 3. More details can be seen in the Supplementary \S1.



\paragraph{Rendering pipeline integration and timing breakdown} 
We integrate our PyTorch-based network inference with \texttt{Falcor}’s framebuffer~\cite{Kallweit22} via CUDA interop. For the local query decoder, we extend \texttt{cudaKDTree}~\cite{wald2023gpu} to accelerate k-nearest neighbor search. We report average runtime on our test set: (1) The \textit{Light Transport Encoder} takes 208 ms which is dominated by the transformer; however, as our method is view-independent, this cost is incurred only once per scene update and is effectively amortized across frames. (2) The rendering phase \textit{Local Query Decoder} requires 368\,ms for generating a $512 \times 512$ resolution image on a single NVIDIA RTX 5880 Ada GPU, which includes $\sim$25--36\,ms for GPU KNN search with the rest spent on Python-based attention. Although we have not achieved real-time framerates yet, this suggests that fusing the attention layers into a custom CUDA kernel or applying model distillation could yield significant speedups in future work.

\paragraph{Loss function} We optimize the entire network end-to-end to minimize the relative L2 loss in log space, applied to both estimated irradiance and radiance~\cite{muller2021radiancecache}:

\begin{equation}
    \mathcal{L} = \frac{1}{N} \sum_{j} \left( \frac{\log(\hat{\mathbf{I}}_j + 1) - \log(\mathbf{I}_j^{\text{gt}} + 1)}{\log(\hat{\mathbf{I}}_j + 1) + \epsilon} \right)^2
    \label{eq:loss_function}
\end{equation}
\noindent where $\hat{\mathbf{I}}_j$ and $\mathbf{I}_j^{\text{gt}}$ denote the predicted and ground truth values for the $j$-th query point, respectively, and $N$ represents the total number of query points in the batch.






\section{Experiments and analysis}
We primarily validate our framework on Diffuse Global Illumination (\S\ref{sec:diffuse}), utilizing smooth irradiance prediction as a benchmark for cross-scene generalization. Conceptually, this serves as a generalizable neural analogue to irradiance caching~\cite{ward1992irradiance}, replacing interpolation with robust learned aggregation.

Subsequently, mirroring the evolution toward radiance caching \cite{jarosz2008radiance}, we demonstrate versatility (\S\ref{sec:glossy}) by re-targeting our encoder to predict spatial-directional radiance fields for glossy materials, achieved by simply exchanging the decoder and applying limited fine-tuning. We conclude with a critical analysis of scalability (\S\ref{sec:scalability}) and detailed ablation studies (\S\ref{sec:model_ablation}).

\subsection{Experimental Setup \& Baselines}
We compare against representatives from three distinct paradigms to cover the spectrum of neural rendering, thereby complementing recent works~\cite{zeng2025renderformer} with a broader evaluation scope.

\textsc{Path tracing} provides our physically accurate ground truth. We use 64 spp path tracing as a quality baseline, indicative of a real-time sampling budget. \emph{We emphasize that this work does not seek to displace standard path tracing or production-standard PT plus denoising combinations}, but rather to explore whether pure data priors can directly predict global illumination from 3D scene configurations without sampled illumination cues.

\textsc{Deep shading} \cite{nalbach2017deepshading}.
This method predicts GI using screen-space G-buffer attributes and direct lighting. We overfit their model on our test views using ground-truth direct lighting to match their setup. However, unlike our 3D approach, it operates purely in 2D, making it inherently view-dependent and unaware of off-screen geometry. Furthermore, our pipeline learns from raw light sources without requiring rasterized cues. 

\textsc{\citet{hermosilla2019deep}.} This work is pioneering in 3D light transport learning and close to our setup, 
but limited to single objects without textures. To adapt it to our 
dataset, we upgraded the architecture with target-count sampling (for consistent point density), $8\times$ larger feature dimensions, and memory-efficient PointConv~\cite{wu2019pointconv}. Despite this, the model struggled to generalize, requiring per-scene fine-tuning to achieve reasonable results.

\textsc{RenderFormer}~\cite{zeng2025renderformer}, a more recent transformer approach. We detail its scalability constraints in \S\ref{sec:scalability}.

\subsection{Generalizable Diffuse Global Illumination}
\label{sec:diffuse}
\begin{figure*}[!htbp]
    \centering
    \includegraphics[width=1.0\textwidth]{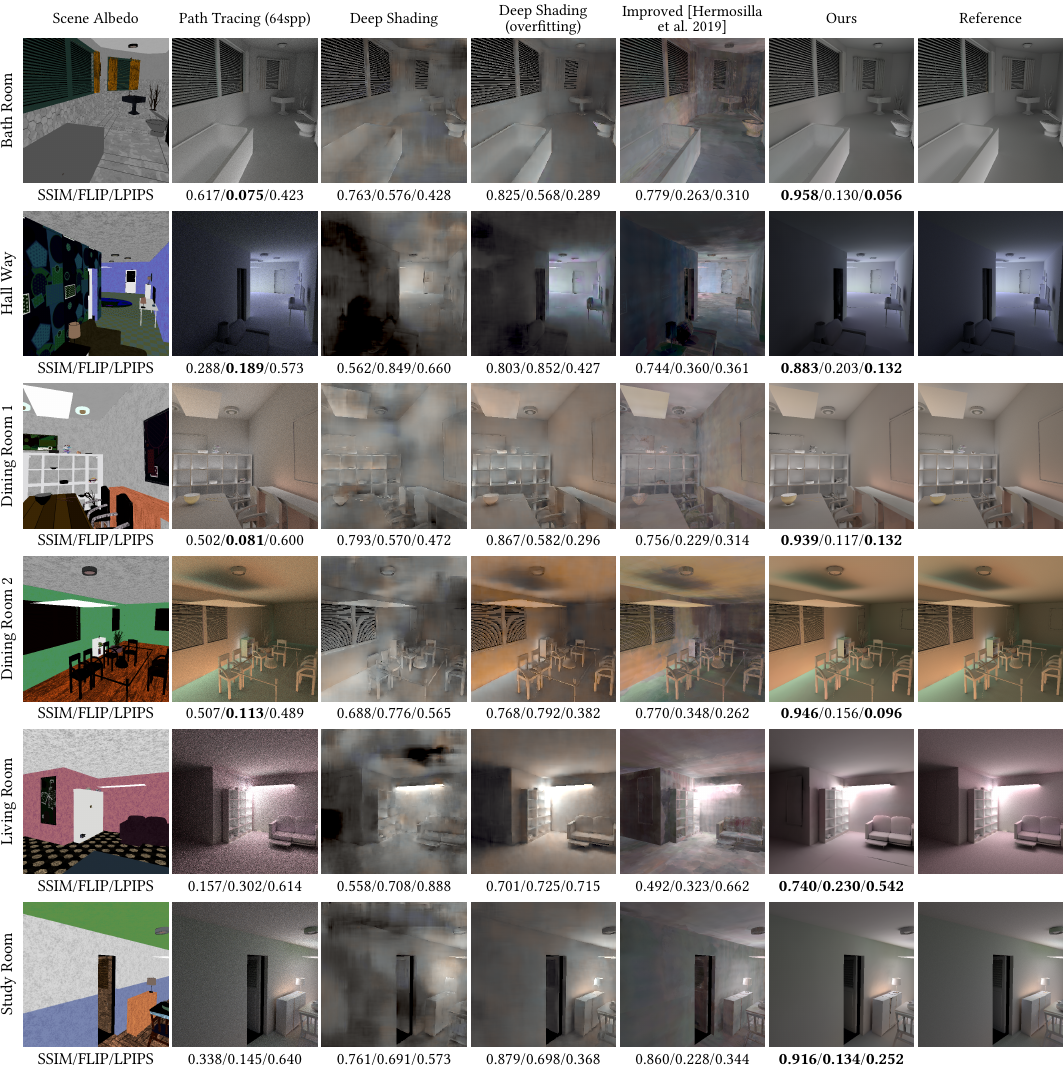} 
    \caption{Irradiance prediction comparison with baseline representatives across six diverse floor-plans under varying illumination conditions. We primarily highlight the substantial qualitative visual improvements while reporting SSIM ($\uparrow$), FLIP ($\downarrow$)~\cite{Andersson2020}, and LPIPS ($\downarrow$)~\cite{zhang2018perceptual} as per-scene perceptual quality references. We evaluate our model on a held-out test set to demonstrate its generalization capability. Notably, no per-scene training is performed, unlike prior neural global illumination approaches. For Deep Shading, which fails to generalize to new scenes, we include the test scenes in their training set to show their \textbf{overfitting} results here. We extend the original implementation by \citet{hermosilla2019deep} to handle our more complex scenes, as their original setup is limited to simple object-level inputs and performs poorly on our dataset (see \S\ref{sec:diffuse}). The first column shows the corresponding albedo to provide readers with context for the color of the indirect lighting, while we avoid showing albedo-modulated images to maintain a clear focus on the quality of the predicted irradiance without distraction. We note that path tracing is included as a real-time-budget rendering reference (\S\ref{sec:diffuse}) rather than a head-to-head competitor; its error profile is dominated by uniform high-frequency noise, which SSIM and LPIPS reflect more directly while FLIP can score favorably.}
    \label{fig:main_comparison}
\end{figure*}

We evaluate on our complex indoor dataset (excluding ~\cite{zeng2025renderformer} due to OOM on these complex scenes, see \S\ref{sec:scalability}).
We present qualitative results on six diverse floor plans in \Cref{fig:main_comparison}, visualizing irradiance without texture modulation to highlight global light transport. \Cref{fig:full_render} shows fully rendered images with texture and direct lighting. In all other cases, visible colors originate purely from indirect lighting. 

The \textsc{hallway} and \textsc{living room} setups are especially challenging for path tracing, with the light source placed in a distant corner so parts of the room are occluded from the main illuminated area. 
Our method consistently outperforms neural baselines and faithfully captures color bleeding from multi-bounce paths and soft shadows. 

Minor artifacts remain: light leaks at the toilet base (\textsc{bath room}) and beneath the bowl (\textsc{dining room 1}), and slight color shifts in \textsc{living room}.
Quantitative results on the test set are provided in \cref{tab:quant_metrics}.
Extended baseline analysis are provided in Supplementary \S3. Furthermore, we demonstrate view consistency and various scene editing applications in Supplementary \S4.

\begin{table}[!htbp]
\centering
\caption{Quantitative comparison (MSE/SSIM) on 105 test scenes. Note that Deep Shading is overfitted to evaluated views due to poor generalization. Our visual quality significantly outperforms all baselines.}
\renewcommand{\arraystretch}{1.1}
\setlength{\tabcolsep}{4pt} 
\begin{tabular}{lcccc}
\toprule
\makecell{\textbf{Metric}} &
\makecell{Deep Shading\\(overfitting)} &
\makecell{Improved \\ \lbrack Hermosilla et al.\rbrack} &
\makecell{Path Tracing} &
\makecell{Ours} \\
\midrule
MSE $\downarrow$     & 0.071 & 0.171 & 0.051 &  0.048 \\
SSIM $\uparrow$      & 0.882 & 0.775  & 0.425 & 0.912  \\
\bottomrule
\end{tabular}
\label{tab:quant_metrics}
\end{table}

\begin{figure*}[!t]
\centering
\setlength{\tabcolsep}{0pt}
\renewcommand{\arraystretch}{1.0}
\begin{tabular}{@{}%
  >{\centering\arraybackslash}m{0.165\textwidth}!{\hspace{2pt}}%
  >{\centering\arraybackslash}m{0.165\textwidth}!{\hspace{2pt}}%
  >{\centering\arraybackslash}m{0.165\textwidth}!{\hspace{2pt}}%
  >{\centering\arraybackslash}m{0.165\textwidth}!{\hspace{2pt}}%
  >{\centering\arraybackslash}m{0.165\textwidth}!{\hspace{2pt}}%
  >{\centering\arraybackslash}m{0.165\textwidth}@{}}
\includegraphics[width=\linewidth]{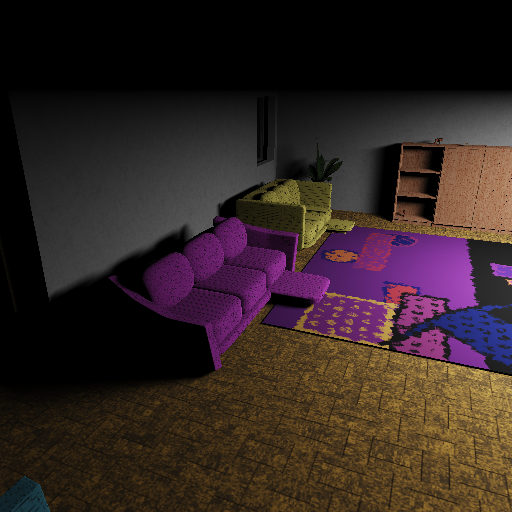} &
\includegraphics[width=\linewidth]{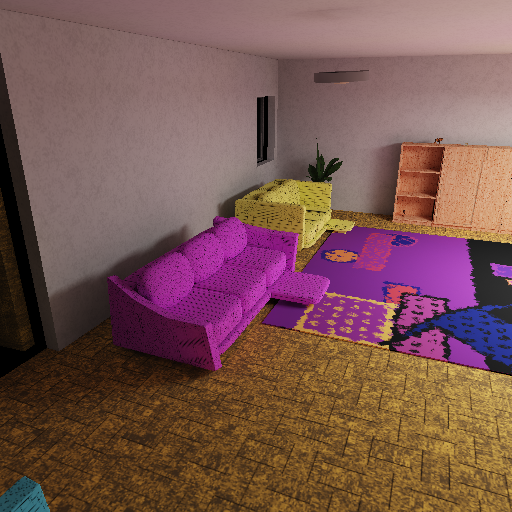} &
\includegraphics[width=\linewidth]{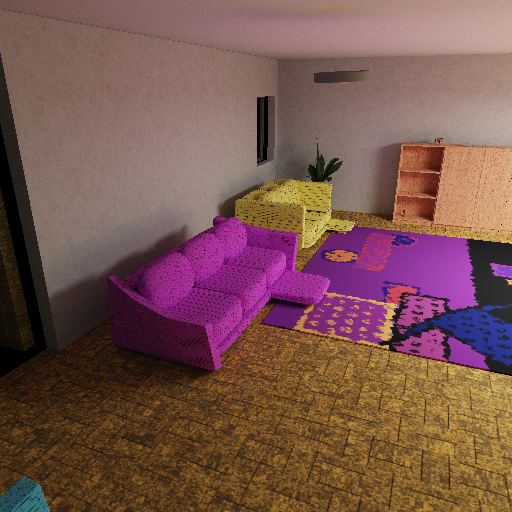} &
\includegraphics[width=\linewidth]{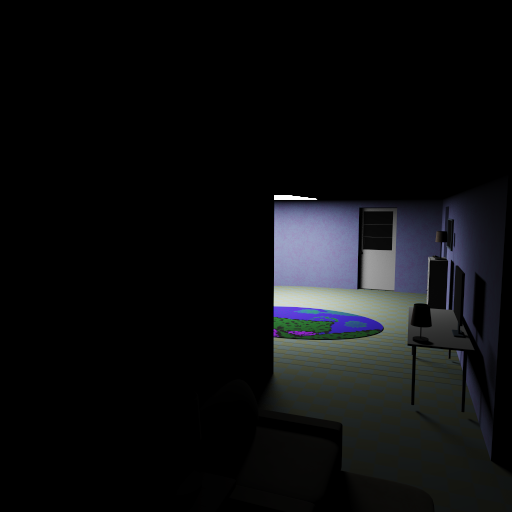} &
\includegraphics[width=\linewidth]{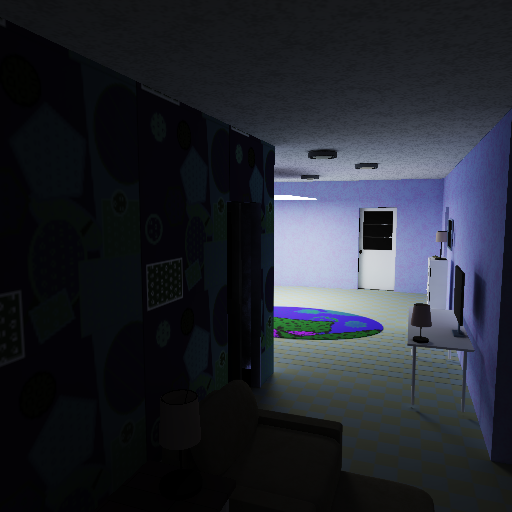} &
\includegraphics[width=\linewidth]{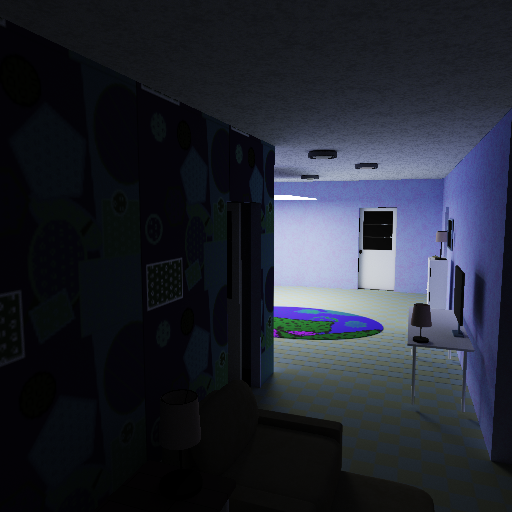} \\
\includegraphics[width=\linewidth]{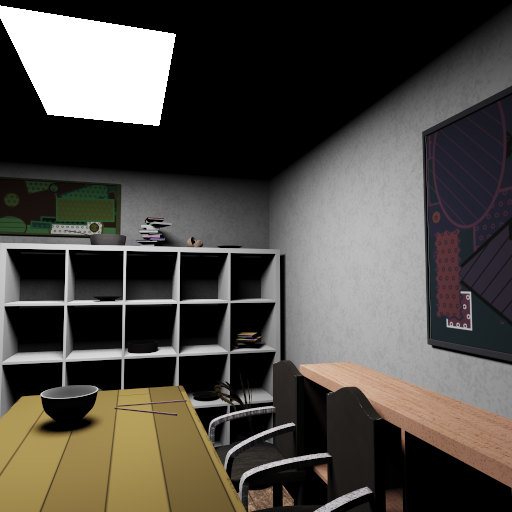} &
\includegraphics[width=\linewidth]{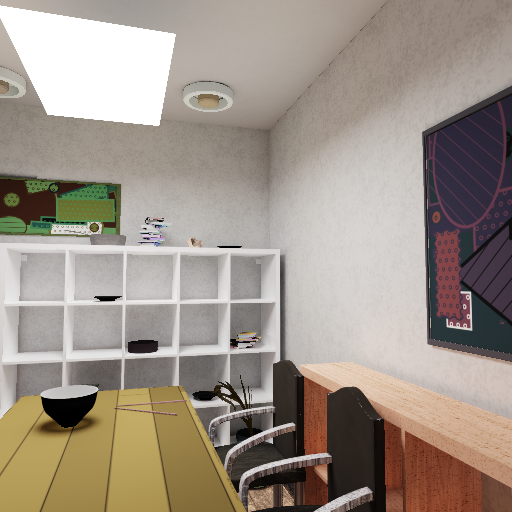} &
\includegraphics[width=\linewidth]{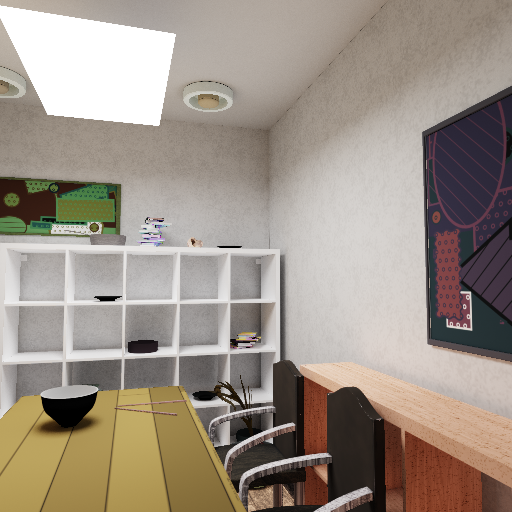} &
\includegraphics[width=\linewidth]{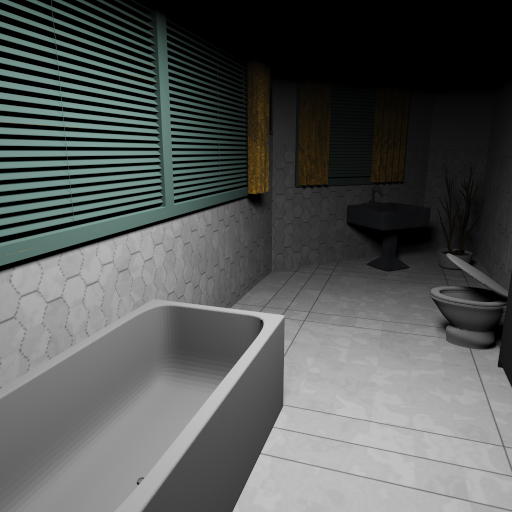} &
\includegraphics[width=\linewidth]{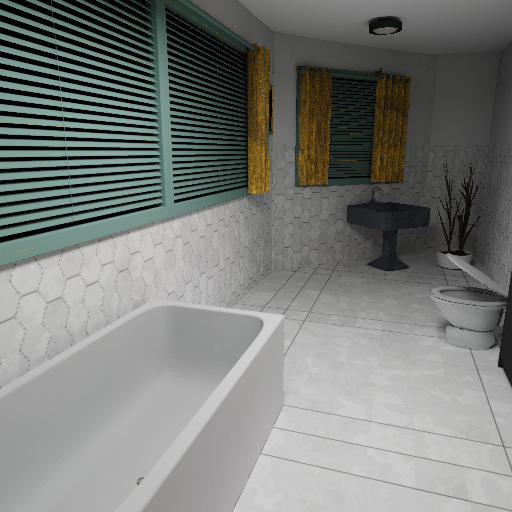} &
\includegraphics[width=\linewidth]{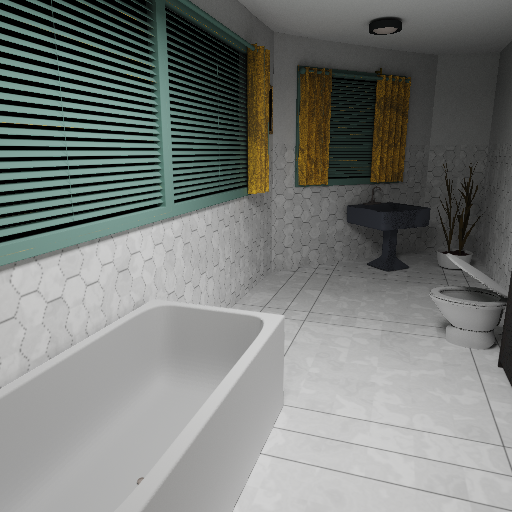} \\
\footnotesize \sffamily \shortstack{Direct lighting} &
\footnotesize \sffamily \shortstack{+ Our predicted G} &
\footnotesize \sffamily \shortstack{Path traced reference} &
\footnotesize \sffamily \shortstack{Direct lighting} &
\footnotesize \sffamily \shortstack{+ Our predicted GI} &
\footnotesize \sffamily \shortstack{Path traced reference} \\
\end{tabular}

\caption{Full renderings with direct illumination and our predicted global illumination for diverse floorplans, lighting conditions and camera views.}
\label{fig:full_render}
\end{figure*}


\subsection{Versatility: Spatial-Directional Radiance Fields}
\label{sec:glossy}

While irradiance evaluates our core capability to encode global interactions and isolates the scalability challenges central to this work, the architecture itself is not tied to diffuse transport. We re-purpose our pre-trained encoder to predict spatial-directional radiance fields, which can be directly integrated with glossy BRDF models, demonstrating versatility to view-dependent light transport.

\begin{figure}[!htbp]
    \centering
    \includegraphics[width=\linewidth]{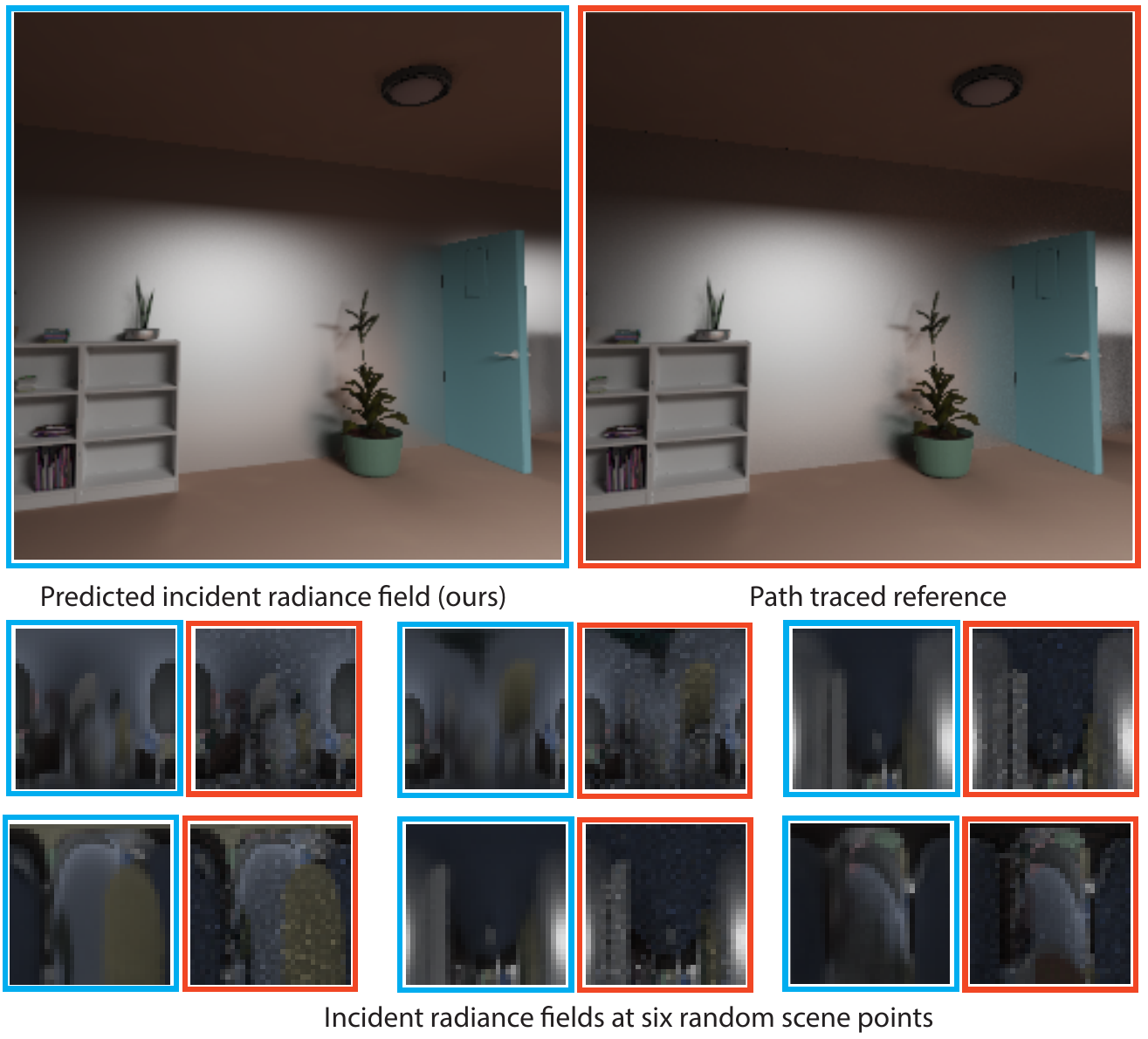}
    \captionof{figure}{Rendering comparison for a mixed scene with diffuse and glossy materials. The wall and plant container use a conductor BRDF modeled with a Trowbridge–Reitz (GGX) microfacet distribution and a roughness of 0.1. For glossy surfaces, our Directional Radiance Decoder directly predicts incident radiance given a shading point and any incoming direction. Global illumination (top left) uses BRDF integrated with our predicted directional radiance field, and the top right shows a path traced reference (1024 directional bins per point and 64 samples per bin). The $32\times32$ slices visualize incident radiance fields at six randomly selected shading points, compared against path-traced reference over 1024 hemispherical directions.}
    \label{fig:glossy_plots}
\end{figure}

We construct a focused dataset for these preliminary experiments by moving objects within a living room floorplan and assigning glossy materials using the Trowbridge–Reitz (GGX) microfacet distribution (see \Cref{fig:glossy_plots}). We add roughness as a material attribute and use a dedicated Radiance Decoder, taking sampled hemispherical directions to predict directional incident radiance. The ground truth is generated via path tracing using cosine-weighted sampling, discretized into 1024 ($32\times32$) directional bins with 64 samples per bin. 
We initialize the encoder from our pre-trained irradiance model and fine-tune end-to-end with the new Radiance Decoder, reusing the same encoder backbone for this view-dependent task. This indicates that the learned representation is transferable across rendering tasks rather than tied to its original training objective. Incident radiance field fitting results are visualized in \Cref{fig:glossy_plots}. 

Our current directional histogram implementation is more expressive than parametric bases such as spherical harmonics for high-frequency glossy structure (see Supp.\ \S4) but remains a discrete, low-resolution representation. Combining the same neural conditioning with compact directional bases such as Spherical Gaussians or Anisotropic Spherical Gaussians, where the network predicts basis parameters rather than per-bin values, is a natural extension we leave as future work. We also explore applications for bootstrapping path guiding in Supplementary \S4.

\subsection{Scalability and Design Discussion}
\label{sec:scalability}

\begin{table}[!htbp]
\centering
\caption{Structural comparison with RenderFormer~\cite{zeng2025renderformer}. $M$ denotes scene primitives (triangles/points); $N$ denotes query points/ray bundles. Ours achieves linear scalability with significantly fewer parameters. The constant \textit{C} is explained in the footnote.}
\small 
\setlength{\tabcolsep}{5pt} 
\begin{tabularx}{\linewidth}{l c c}
\toprule
\textbf{Feature} & \textbf{\cite{zeng2025renderformer}} & \textbf{Ours} \\
\midrule
\multicolumn{3}{l}{\textbf{1. Model Architecture}} \\
\hspace{3mm}\textit{Encoder Complexity} & Quadratic $\mathcal{O}(M^2)$ & {Linear $\mathcal{O}(\mathit{C}M)$\footnotemark, $\mathit{C}$ = 256} \\

\hspace{3mm}\textit{Decoder Complexity} & $\mathcal{O}(MN + N^2)$ & { $\mathcal{O}(\mathit{K}N)$, $\mathit{K}$ = 32} \\
 & Global attention & Local attention \\
\hspace{3mm}\textit{Model Parameters} & $ 205$M / $483$M & {42 M} \\
\midrule
\multicolumn{3}{l}{\textbf{2. Scene Representation}} \\
\hspace{3mm}\textit{Input} & Mesh Vertices & Sampled Point Cloud\\
\hspace{3mm}\textit{Material Support} & Untextured & {Textured} \\
\midrule
\multicolumn{3}{l}{\textbf{3. Training Supervision}} \\
\hspace{3mm}\textit{Target} & 2D Images & {Sampled 3D query points} \\
\hspace{3mm}\textit{Consistency} & View-dependent & {View-independent} \\
\hspace{3mm}  & Resolution-fixed & {Resolution-independent}\\
\bottomrule
\end{tabularx}%
\label{tab:method_comparison}
\end{table}
\footnotetext{\textbf{Derivation:} The cost is derived as $N_{\text{patches}} \times \text{Cost}_{\text{patch}}$. With patch size $=1024$ and downsampling factor $(1/2)^2$ in our Nearest Neighbor Embedding (\S\ref{sec:global_encoding}), we have: 
    $\frac{M/4}{1024} \times 1024^2 = 256 M$. }

We analyze the structural divergence between our approach and the vertex-based RenderFormer~\cite{zeng2025renderformer} in Table~\ref{tab:method_comparison}. While both use transformers, these methods occupy distinct points in the design space of the rendering problem and optimize for different goals: theirs focuses on capturing high-frequency transport on object-centric meshes, whereas our design prioritizes the scalability required for general environments, offering support for textures, flexible resolutions and views. As such, the two approaches are complementary rather than competing.

(1) \textit{The Scalability Barrier}.\footnote{Recall that $M$ denotes the number of scene primitives (points/triangles) and $N$ denotes the number of queries (query points/pixels/ray bundles). Scalability here mainly refers to computational scalability -- the ability of an architecture to handle increasing scene complexity (primitives, resolution, etc.) without a prohibitive increase in resource consumption, specifically GPU memory.} RenderFormer employs global attention to model complex long-range interactions. While effective for object-level effects, this has two limitations. First, this incurs quadratic costs across all dimensions: geometry self-attention $\mathcal{O}(M^2)$, ray self-attention $\mathcal{O}(N^2)$, and cross-attention $\mathcal{O}(MN)$. Second, this joint attention mechanism tightly couples scene complexity ($M$) and rendering resolution ($N$), creating a ``\textbf{zero-sum}'' competition for the GPU memory. Consequently, its model's learning capacity is constrained by relatively simple training meshes ($\approx 4k$ triangles) and low resolutions. \footnote{We note that the $\approx 4k$-triangle figure refers to the training mesh budget; RenderFormer can be invoked on higher-poly meshes at inference, but at a quality cost reflecting the training distribution.}
This bottleneck is illustrated in Figure~\ref{fig:scalability_renderformer}. In contrast, we prioritize linear scalability and decouple the per-query computational cost from the global scene complexity. By leveraging the point representation discussed below, a linear-complexity encoder and a constant local decoder ($K=32$), we scale to dense indoor scenes with millions of triangles using only \textbf{42M} parameters, while supporting arbitrary rendering resolutions.

\begin{figure}[!t]
\centering
\setlength{\tabcolsep}{0pt}
\renewcommand{\arraystretch}{1.0}
\begin{tabular}{@{}cccc@{}}
\includegraphics[width=0.25\linewidth]{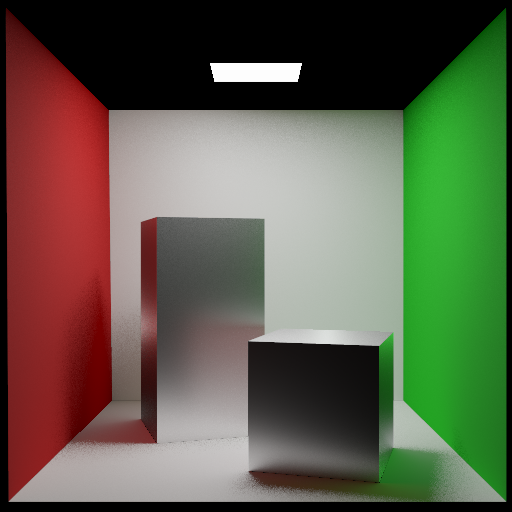} &
\begin{tikzpicture}[inner sep=0pt]
  \node[anchor=south west] (img) {\includegraphics[width=0.25\linewidth]{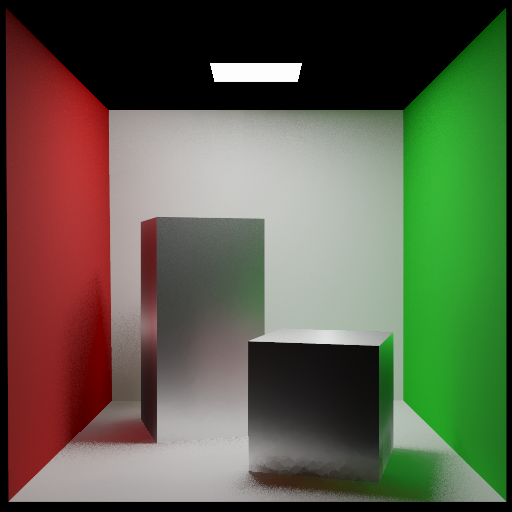}};
  \node[anchor=north east, inner sep=0pt, draw=white, line width=0.4pt]
    at ([xshift=-1pt, yshift=-1pt]img.north east)
    {\includegraphics[width=0.085\linewidth]{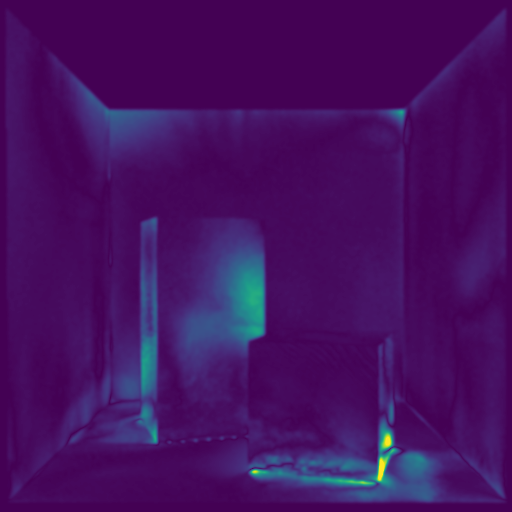}};
\end{tikzpicture} &
\includegraphics[width=0.25\linewidth]{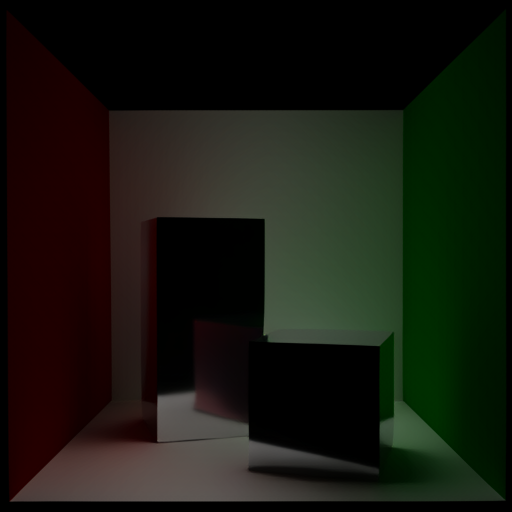} &
\includegraphics[width=0.25\linewidth]{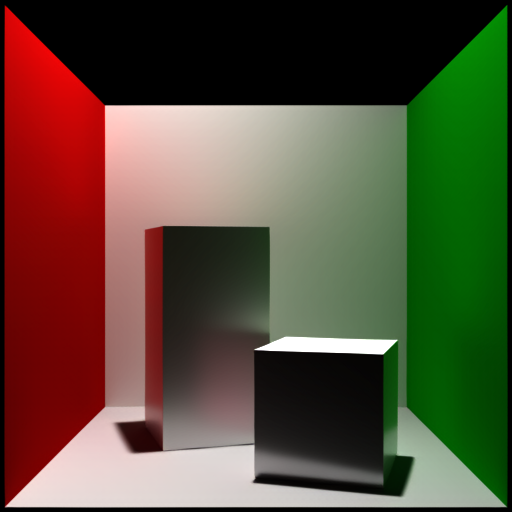} \\
{\footnotesize\sffamily Reference} &
{\footnotesize\sffamily Ours ($5{\times}$ error inset)} &
{\footnotesize\sffamily \shortstack{RenderFormer\\[-2pt]\scriptsize(Cbox, RF-adapted)}} &
{\footnotesize\sffamily \shortstack{RenderFormer\\[-2pt]\scriptsize(their own example)}} \\
\end{tabular}\vspace*{-6pt}
\caption{Cross-method comparison on a Cornell Box, the classical small-mesh setting where RenderFormer is designed to operate. Left to right: path-traced reference; our radiance-decoder prediction ($5\times$ error inset \cite{Andersson2020}); RenderFormer's rendering of the same Cornell Box adapted to its expected input scene assumption, including mesh tessellation, scene normalization, and point-light approximation (see Supplementary \S3); and and an official result from the original RenderFormer publication. The latter is included to ensure the comparison remains faithful to their official benchmarks and to provide a reference within the model’s intended training distribution. Our model is fine-tuned for one day on a small low-poly subset.}
\label{fig:comparison_renderformer}\vspace*{-6pt}
\end{figure}

(2) \textit{Scene Representation.} Vertex-based transformers couple geometry and appearance. In RenderFormer, material attributes are bound to vertices, so capturing high-frequency texture details requires increasing vertex density (tessellation) or encoding attributes at triangle level as in \citet{zeng2025renderformer}, which in turn triggers the memory bottleneck. 
Our point-based approach decouples these factors. We can sample points densely from textured surfaces to faithfully model complex indoor environments with spatially varying materials. This also unifies emissive and non-emissive geometries, making the architecture more adaptable to various light sources. As a quick test to demonstrate that our pipeline can also handle the low-poly regime where RenderFormer natively operates, we additionally fine-tune our model on a small low-poly subset to provide a functional comparison (see Figure~\ref{fig:comparison_renderformer}).

(3) \textit{Supervision.} Training objectives dictate tradeoffs. RenderFormer minimizes loss in 2D image space; its 2D supervision enables it to faithfully model view-dependent caustics and specular highlights, namely, the loss directly sees what the camera sees. However, this approach risks overfitting to constrained viewing frustum and fixed resolution. By supervising directly in 3D world space (\S\ref{sec:training_scheme}), we trade off the ability to model high-frequency view-dependent effects for a representation that is inherently view-independent and resolution-agnostic. Both strategies represent valid solutions to distinct optimization problems. See the Supplementary material for further discussion, particularly Supp. Table 4.


\subsection{Ablation Studies and Analysis}
\label{sec:model_ablation}
\begin{figure}[!htbp]
    \centering
    \setlength{\tabcolsep}{0pt}
    \renewcommand{\arraystretch}{1.0}
    \begin{tabular}{@{}%
      >{\centering\arraybackslash}m{0.25\linewidth}%
      >{\centering\arraybackslash}m{0.25\linewidth}%
      >{\centering\arraybackslash}m{0.25\linewidth}%
      >{\centering\arraybackslash}m{0.25\linewidth}%
    @{}}
      \includegraphics[width=\linewidth]{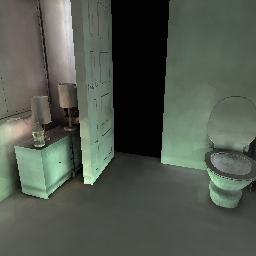} &
      \includegraphics[width=\linewidth]{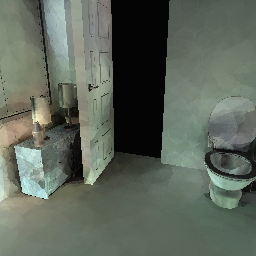} &
      \includegraphics[width=\linewidth]{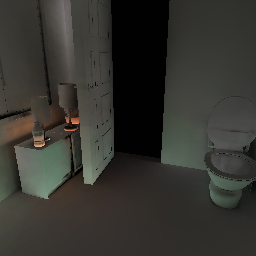} &
      \includegraphics[width=\linewidth]{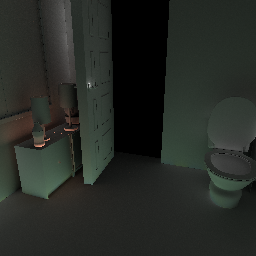} \\
      \footnotesize \shortstack{Vanilla\\Transformer} &
      \footnotesize \shortstack{+ Light Transport\\Encoder (ours)} &
      \footnotesize \shortstack{+ Local Query\\Decoder (ours)} &
      \footnotesize Reference \\
    \end{tabular}

    \captionof{figure}{Qualitative results for ablating different components of the model. This ablation also serves as a \textbf{roadmap} of our design and experimental journey. \textbf{Left:} Global vanilla Transformer encoder, an intuitive starting point, but quadratic attention makes it non-scalable with point count and hard to train. \textbf{Middle:} Our Light Transport Encoder (\S\ref{sec:global_encoding}) with naive KNN aggregation: highly scalable and fast, but with a noticeable quality drop. \textbf{Right:} Adding the Local Query Decoder (\S\ref{sec:local_query_decoding}) completes the design and recovers quality, achieving the best overall performance.}

    \label{fig:ablation_qual}

    \definecolor{roadmapBlue}{RGB}{0, 114, 178}
    \definecolor{roadmapOrange}{RGB}{230, 159, 0}

    \begin{tikzpicture}
      \pgfplotsset{set layers}

      \begin{axis}[
        width=\linewidth, height=3.8cm,
        font=\sffamily,
        symbolic x coords={Vanilla, LTE, LQD},
        xtick=data,
        xticklabels={Vanilla Transformer, + Light Transport Encoder, + Local Query Decoder (ours)},
        xticklabel style={text width=0.30\linewidth, align=center, font=\footnotesize\sffamily},
        ylabel={MSE $\downarrow$},
        ylabel style={font=\footnotesize\sffamily, text=roadmapBlue!75!black},
        yticklabel style={font=\footnotesize\sffamily, text=roadmapBlue!75!black},
        axis y line*=left,
        axis x line*=bottom,
        grid=major,
        grid style={dashed, gray!25},
        ymin=0.040, ymax=0.098,
        ytick={0.04, 0.05, 0.06, 0.07, 0.08, 0.09},
        axis line style={gray!50},
        tick style={gray!50},
      ]
        \addplot+[line width=1.3pt, color=roadmapBlue, densely dashed, mark=*, mark size=2.2pt,
          mark options={fill=roadmapBlue, draw=roadmapBlue, line width=1pt}]
          coordinates {(Vanilla,0.068) (LTE,0.084) (LQD,0.048)}
          node[pos=0,   above=4pt, font=\footnotesize] {0.068}
          node[pos=0.5, above=4pt, font=\footnotesize] {0.084}
          node[pos=1,   above=4pt, font=\footnotesize] {\textbf{0.048}};
      \end{axis}

      \begin{axis}[
        width=\linewidth, height=3.8cm,
        font=\sffamily,
        symbolic x coords={Vanilla, LTE, LQD},
        xtick=data,
        xticklabel=none,
        ymin=0.855, ymax=0.920,
        yticklabel style={/pgf/number format/fixed, /pgf/number format/precision=3,
                          font=\footnotesize\sffamily, text=roadmapOrange!80!black},
        ylabel={SSIM $\uparrow$},
        ylabel style={font=\footnotesize\sffamily, text=roadmapOrange!80!black},
        axis y line*=right,
        axis x line=none,
        axis line style={gray!50},
        tick style={gray!50},
      ]
        \addplot+[line width=1.3pt, color=roadmapOrange, mark=square*, mark size=2.2pt,
          mark options={fill=roadmapOrange, draw=roadmapOrange, line width=1pt}]
          coordinates {(Vanilla,0.885) (LTE,0.870) (LQD,0.908)}
          node[pos=0,   below=4pt, font=\footnotesize] {0.885}
          node[pos=0.5, below=4pt, font=\footnotesize] {0.870}
          node[pos=1,   below=4pt, font=\footnotesize] {\textbf{0.908}};
      \end{axis}
    \end{tikzpicture}

    \captionof{figure}{Quantitative complement (100 random test scenes) aligned with the same ablation roadmap. MSE (left; lower is better) and SSIM (right; higher is better) are reported for the same sequence of design choices.}
    \label{fig:ablation_quant}
\end{figure}

We validate our architecture through a progressive ablation study, detailed in Figures~\ref{fig:ablation_qual} and~\ref{fig:ablation_quant}. 
The baseline \textsc{Vanilla} model uses a basic global transformer as an encoder. For each query point, it retrieves neighboring scene points via k-nearest neighbors (KNN), followed by simple pooling and a multi-layer perceptron as the decoder. 
Despite promising results, 
training suffers from quadratic complexity of global attention, making it inefficient for high-resolution point clouds.

To address this, we substitute the global transformer with our Light Transport Encoder. While this intermediate step (using naïve KNN decoding) incurs a slight performance drop, it establishes the necessary scalability. Our Full Model completes the pipeline by incorporating the Local Query Decoder; this learnable aggregation recovers the lost fidelity and achieves the best overall performance. This ablation study reinforces that each component contributes to the final model's effectiveness. Beyond these core components, a comprehensive suite of additional ablations on further design choices (e.g. scene and query point sampling, hyper-parameters, etc) is provided in Supplementary~\S2.

\section{Limitations}
\label{sec:failurecase}
\begin{figure}[!htbp]
    \centering
    \includegraphics[width=\linewidth]{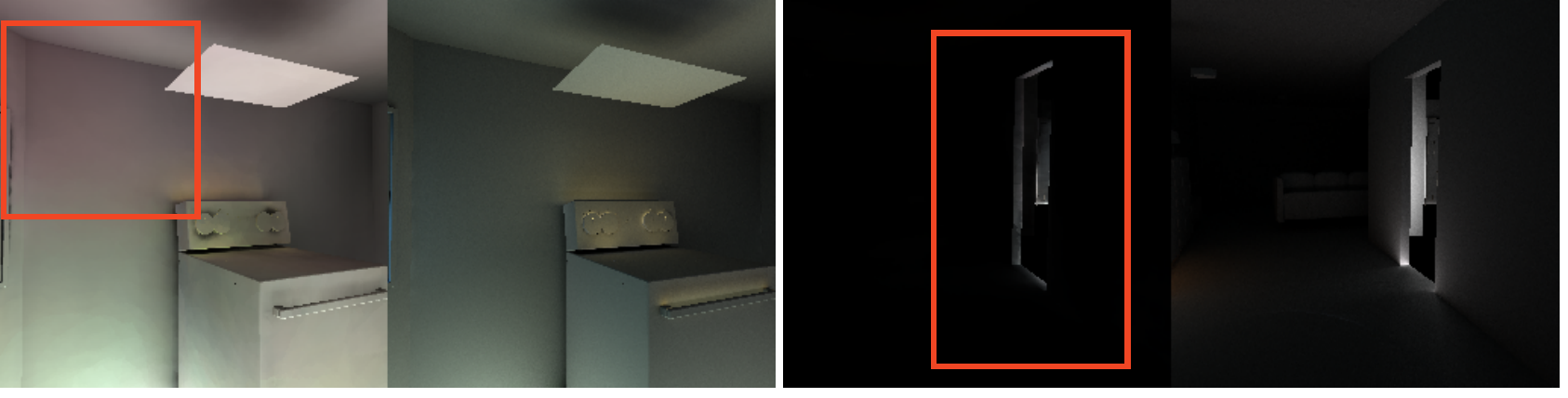}
    \captionof{figure}{Two failure cases discussed in \S\ref{sec:failurecase}: (\textbf{left}) color shifting under underrepresented texture tones; (\textbf{right}) ``ajar'' lighting where the source is in an adjacent room.}
    \label{fig:failure_cases}
\end{figure}

Our method shares several challenges common to regression-based learning approaches. We observe occasional color shifts in scenes containing underrepresented texture tones, such as the \textsc{Hallway} and \textsc{Study room} shown in \Cref{fig:main_comparison}. We also note degraded predictions when the primary light source is positioned in an adjacent room (\Cref{fig:failure_cases}). These out-of-distribution shifts could be mitigated through targeted dataset augmentation including outdoor environments, or through limited fine-tuning on specific new domains.

Furthermore, visual artifacts can arise from occluded points. For example, points sampled beneath geometry, such as under a carpet in the 5th scene of \Cref{fig:main_comparison}, can occasionally bleed dark latent codes into visible regions. This issue can be resolved by increasing the neighbor count $k$ during decoding or by pruning fully occluded points during the initial sampling stage.

Consistent with other learning-based global illumination methods, our model may exhibit flickering in dynamic sequences without explicit temporal constraints. It also shows reduced fidelity on high-frequency specular reflections because the current encoder is optimized primarily for diffuse-dominated transport. Finally, while inference latency remains a factor, it can be improved via distillation or embedding pruning. We consider this optimization secondary to our primary goal, which is exploring pure data priors for light transport and architectural scalability.

\section{Conclusion and future work}

We have presented a scalable 3D light transport embedding that effectively breaks the memory bottleneck of prior neural rendering methods. By replacing quadratic global attention with a linear-complexity transformer on point clouds and employing independent query point-wise local decoding, our framework enables the prediction of global illumination for indoor environments with millions of primitives. Crucially, this efficiency unlocks robust cross-scene generalization. Beyond scalability, we demonstrated the versatility of this generalizable embedding. Our shared encoder successfully adapts from predicting diffuse GI to modeling view-dependent glossy radiance fields. Our 3D supervision ensures view consistency and resolution independence, overcoming the temporal instability often seen in image-space baselines. Our method highlights the potential of tensor cores as an alternative to ray tracing cores, complementing rather than replacing classical pipelines.

Future work includes incorporating participating media and outdoor scenes to extend the spectrum of light transport effects, alongside more sample-efficient training mechanisms to ensure sustainable scaling. Because our pipeline is differentiable by construction, it can serve as a foundational building block for inverse rendering and scene reconstruction. Additionally, incremental embedding updates that re-encode only regions affected by scene changes would extend our amortization beyond camera-only motion to truly dynamic content. Looking ahead, a shared challenge for all efforts in this direction is finding mechanisms that scale sustainably across both the geometric scene space and the high-dimensional rendering parameter space to capture the full complexity of light transport. A careful and evolving balance between data-driven priors and physical laws remains essential for the future of neural rendering.



\begin{acks} 
This work was supported in part by NSF grants 2212085, 2105806, 2100237, 2120019, 2127544, and ONR grant N00014-23-1-2526.  We also acknowledge support from Apple, gifts from Adobe, Google, Activision, and Qualcomm, the Ronald L. Graham Chair and the UC San Diego Center for Visual Computing.  Ramamoorthi acknowledges a part-time appointment at NVIDIA. The authors would like to thank Aaron Lefohn and Chris Wyman for their support; Matt Pharr for valuable input along the way; Thomas Akenine-Möller, Jacob Munkberg, Jon Hasselgren and Zian Wang for their suggestions regarding the evaluation, dataset and training clusters; Miloš Hašan and Benedikt Bitterli for valuable discussions; and to anonymous reviewers for constructive comments. Bing owes particular thanks to Alex Trevithick for pointers to procedural scene generation and many helpful discussions; Zimo Wang and Nithin Raghavan for generously sharing cluster training quotas; Yang Zhou for proofreading and comments; and her fellow interns for helpful discussions in the early stages. 

\end{acks}

\bibliographystyle{ACM-Reference-Format}
\bibliography{paper}
\appendix


\section{Implementation details}
\label{sec:implementation}
\subsection{Architecture and training configurations}
While the overall architectural design is described in \S3 of main paper, we provide the specific layer configurations, channel dimensions and training settings in Table \ref{tab:arch_details}. 
\begin{table}[h]
\centering
\caption{Summary of architectural specifications and training settings.}
\small
\begin{tabular}{l|l c}
\toprule
\textbf{Category} & \textbf{Parameter} & \textbf{Value} \\
\midrule
\textbf{Model}
  & Pre-projection Dim & 128 \\
  & Nearest-neighbor embedding K & 32 \\
 & Encoder Backbone & PTV3 \\
 & \textit{Structure} & Enc Depths: $[2,2,2,6,2]$ \\
 & \textit{Channels} & $[128, 128, 128, 256, 256]$ \\
 & \textit{Patch Size} & 1024 \\
 & Encoder Heads & 4 ($D/32$) \\
 & Hidden Dimension ($D$) & 128 \\
 & Decoder Layers & 6 \\
 & Decoder Neighbors ($K$) & 32 \\
 & Feed-Forward Dim & 512 ($4 \times D$) \\
 & Dropout & 0.1 \\
\midrule
\textbf{Training} 
 & Optimizer & AdamW \\
 & Learning Rate & $2 \times 10^{-4}$ \\
 & Weight Decay & $1 \times 10^{-4}$ \\
 & Batch Size & 3 \\
 & Total Steps & 2,000,000 \\
\bottomrule
\end{tabular}
\label{tab:arch_details}
\end{table}

\subsection{Background on Point Transformer V3 and our NNE's receptive field expansion}
\label{supp:ptv3}

We adopt Point Transformer V3 (PTV3)~\cite{wu2024pointtransformerv3} as our encoder backbone for its scalability properties. While earlier point-based transformers~\cite{zhao2021pointtransformer} relied on computationally expensive $k$-nearest-neighbor (KNN) lookups for every attention layer, PTV3 introduces a more efficient serialization-and-patch mechanism.

As illustrated in \Cref{fig:ptv3_intuition}, PTV3 reorders 3D points into a 1D sequence using space-filling curves (e.g., Z-order or Hilbert curves). This sequence is partitioned into non-overlapping patches of a fixed size (we use 1024). Standard self-attention is then computed exclusively within these patches. To ensure cross-neighborhood communication, PTV3 shuffles the serialization order (e.g., swapping between Z-order and Hilbert) across different layers. This design achieves linear complexity, allowing our model to process the millions of points required for high-fidelity light transport.

\paragraph{The spatial coverage limitation}
While PTV3 is highly efficient for semantic tasks (like segmentation), its patch-based attention presents a specific challenge for Global Illumination (GI). In GI, a single point's radiance is often determined by distant light sources or large-scale bounces across a room.

A 1024-point patch of raw scene points typically covers only a small, localized 3D volume. Since attention is restricted to the patch, the model can ``see'' only a thin slice of the environment in a single layer. In standard PTV3, long-range information must propagate slowly through many stacked layers or hierarchical downsampling.
This limitation directly motivates our Nearest-Neighbor Embedding (NNE) (\S3.3 of the main paper). 

\begin{figure}[h!]
\centering
\begin{tikzpicture}[scale=0.46, every node/.style={font=\scriptsize\sffamily}]

\begin{scope}[shift={(0,0)}]
  \foreach \x in {0,1,2,3}
    \foreach \y in {0,1,2,3}
      \fill[gray!70] (\x, \y) circle (0.18);
  \node[align=center] at (1.5, -1.2) {(a) raw\\ scene points};
\end{scope}

\begin{scope}[shift={(5,0)}]
  \draw[blue!70, line width=1.0pt, ->,>=stealth]
    (0,0) -- (0,1) -- (1,1) -- (1,0) -- (2,0) -- (3,0) -- (3,1) -- (2,1)
    -- (2,2) -- (3,2) -- (3,3) -- (2,3) -- (1,3) -- (1,2) -- (0,2) -- (0,3);
  \foreach \x in {0,1,2,3}
    \foreach \y in {0,1,2,3}
      \fill[blue!70] (\x, \y) circle (0.18);
  \node[align=center] at (1.5, -1.2) {(b) space-filling\\ serialization};
\end{scope}

\begin{scope}[shift={(10,0)}]
  \draw[orange, line width=1.2pt] (0,0) -- (0,1) -- (1,1) -- (1,0);
  \fill[orange] (0,0) circle (0.20);
  \fill[orange] (0,1) circle (0.20);
  \fill[orange] (1,1) circle (0.20);
  \fill[orange] (1,0) circle (0.20);
  \draw[orange!60, dashed, line width=0.6pt] (0,0) -- (1,1);
  \draw[orange!60, dashed, line width=0.6pt] (0,1) -- (1,0);

  \draw[green!55!black, line width=1.2pt] (2,0) -- (3,0) -- (3,1) -- (2,1);
  \fill[green!55!black] (2,0) circle (0.20);
  \fill[green!55!black] (3,0) circle (0.20);
  \fill[green!55!black] (3,1) circle (0.20);
  \fill[green!55!black] (2,1) circle (0.20);
  \draw[green!55!black, dashed, line width=0.6pt, opacity=0.6] (2,0) -- (3,1);
  \draw[green!55!black, dashed, line width=0.6pt, opacity=0.6] (3,0) -- (2,1);

  \draw[purple, line width=1.2pt] (2,2) -- (3,2) -- (3,3) -- (2,3);
  \fill[purple] (2,2) circle (0.20);
  \fill[purple] (3,2) circle (0.20);
  \fill[purple] (3,3) circle (0.20);
  \fill[purple] (2,3) circle (0.20);
  \draw[purple!60, dashed, line width=0.6pt] (2,2) -- (3,3);
  \draw[purple!60, dashed, line width=0.6pt] (3,2) -- (2,3);

  \draw[red!80, line width=1.2pt] (1,3) -- (1,2) -- (0,2) -- (0,3);
  \fill[red!80] (1,3) circle (0.20);
  \fill[red!80] (1,2) circle (0.20);
  \fill[red!80] (0,2) circle (0.20);
  \fill[red!80] (0,3) circle (0.20);
  \draw[red!80, dashed, line width=0.6pt, opacity=0.6] (1,3) -- (0,2);
  \draw[red!80, dashed, line width=0.6pt, opacity=0.6] (1,2) -- (0,3);

  \node[align=center] at (1.5, -1.2) {(c) per-patch\\ attention};
\end{scope}

\begin{scope}[shift={(15,0)}]
  \fill[teal!15] (0.5, 0.5) circle (1.3);
  \fill[teal!15] (2.5, 0.5) circle (1.3);
  \fill[teal!15] (2.5, 2.5) circle (1.3);
  \fill[teal!15] (0.5, 2.5) circle (1.3);

  \foreach \x in {0,1,2,3}
    \foreach \y in {0,1,2,3}
      \fill[gray!55] (\x, \y) circle (0.13);

  \fill[teal!75!black] (0.5, 0.5) circle (0.26);
  \fill[teal!75!black] (2.5, 0.5) circle (0.26);
  \fill[teal!75!black] (2.5, 2.5) circle (0.26);
  \fill[teal!75!black] (0.5, 2.5) circle (0.26);

  \node[align=center] at (1.5, -1.2) {(d) NNE tokens\\ + coverage halos};
\end{scope}

\end{tikzpicture}
\caption{Visual sketch of PTV3's serialization-and-patch design and our NNE complement.
(a) Raw scene points.
(b) Points are reordered along a space-filling curve (a Hilbert curve is shown; PTV3 also supports Z-order and transposed variants).
(c) The 1D sequence is split into non-overlapping patches (colored); attention is computed only within each patch.
(d) Our Nearest-Neighbor Embedding (NNE; \S3.3 of the main paper) replaces raw scene points with hierarchically aggregated tokens: two FPS halvings yield $m{=}M/4$ centers (teal), and each token (large teal dot) summarizes a halo (faded teal) of $\sim k^2$ raw points via two stages of $k$-NN aggregation. A patch of 1024 such tokens covers a much larger 3D region than a 1024-raw-point patch in (c) would.}
\label{fig:ptv3_intuition}
\end{figure}

\subsection{Dataset generation details}
\label{supp:dataset}

Our large-scale indoor global illumination dataset is built upon the procedural generation rules from \citet{raistrick2024infinigen}. The dataset comprises 13,900 Blender scenes spanning diverse floorplans and asset collections, including furniture, appliances, and decor.

\paragraph{Geometry refinement} We clean up the geometries to ensure geometric consistency suitable for physically based rendering. This involved:
(1) Filtering out low-quality meshes and degenerate geometry. (2) {De-duplication:} Removing duplicate elements and resolving overlapping exterior shells that could cause light leakage or z-fighting artifacts. (3) All scenes were processed into PBRT format with high-resolution baked textures to preserve material fidelity during export.

\paragraph{Lighting heuristics} We utilized area lights to provide visually pleasing near-field illumination, unifying them with general scene geometries via our point-based representation. To generate realistic configurations automatically, we implemented a heuristic strategy where rectangular lights were instantiated flush with the ceiling to mimic standard fixtures, while enforcing minimum distance constraints from walls and major occluders to prevent unrealistic hotspots or geometric intersections.

\paragraph{Ground truth generation} Each scene is exported to PBRT format, with high-quality triangle meshes and baked high-resolution textures for each asset. The resulting files range from 500 MB to 2 GB. For ground-truth irradiance computation, we sample 2 million query points per scene and trace 1,024 stratified, cosine-weighted rays per point with a maximum depth of five. 
Full dataset generation, including mesh processing, texture baking, and rendering, required approximately 250 CPU/GPU days on our compute cluster.

\paragraph{Glossy subset} To support the versatility experiments in \S5.3, we generated a focused subset of scenes where glossy surfaces are modeled as conductors using the Trowbridge–Reitz (GGX) microfacet distribution~\cite{walter2007microfacet}. This subset utilizes our standard lighting and geometry pipeline but incorporates additional material roughness attributes and directional incident radiance targets. At each query point, we sampled 1,024 hemispherical directions ($32 \times 32$) using cosine-weighted, stratified sampling. We traced paths along each direction to compute incident radiance, resulting in a discretized $32 \times 32$ directional histogram (and associated PDFs) at every point. We note that this dense representation incurs a significant storage footprint; for future scaling to larger datasets, we recommend adopting on-the-fly BRDF importance sampling to mitigate these memory costs for training data.

\paragraph{Training scene normalization}
Light transport is scale-sensitive; phenomena like inverse-square falloff and occlusion depend on absolute physical distances. Therefore, we do not normalize each scene to its own unit cube. Instead, we compute the scene-specific bounding box minimum $\mathbf{b}_{\min}$ and apply a consistent global scale:
\begin{equation*}
    \hat{\mathbf{p}} = (\mathbf{p} - \mathbf{b}_{\min}) / s,
\end{equation*}
where $s$ is a fixed dataset-wide constant representing an upper bound on scene extent. This ensures every training scene fits within $[0,1]^3$ while maintaining a consistent mapping between world-space and normalized distances. The same transform is applied to query points, while normals and albedos remain unchanged. Under this scheme, the encoder sees physical scale consistently across all environments. Scenes substantially larger than $s$ would either require partitioning or a rescaled $s$, while the remainder of the pipeline remains unchanged.

\section{Ablation studies and analysis}
\label{sec:ablation_study}
We have made specific design choices regarding training data generation and network architecture. In this section, we present ablation studies to validate these decisions and analyze the impact of key hyperparameters on model performance.

\paragraph{Resource cost and visual artifacts in the design roadmap}
We provide further details regarding the ablation roadmap (Fig.~11) by quantifying resource consumption and addressing specific visual artifacts. The \textsc{Vanilla} transformer baseline, while inherently expressive due to its dense attention mechanism, is memory-prohibitive at scale: with $M=20\text{k}$ scene points, it exceeds 80\,GB of training VRAM at a batch size of 3, necessitating a batch size of 1 in practice. Our Light Transport Encoder reduces peak training VRAM to $\approx$21\,GB at batch size 3, while the full pipeline (including the Local Query Decoder) utilizes $\approx$22\,GB. All configurations were trained for 100k steps.

Notably, the hexagonal patterns visible in the intermediate stage (+Light Transport Encoder) are Voronoi-like artifacts inherent to point-based representations. Since FPS downsampling produces a blue-noise point distribution, naive KNN aggregation implicitly partitions the 3D space into Voronoi cells, resulting in visible discontinuities at cell boundaries. As shown in the final column, our Local Query Decoder successfully resolves these artifacts through learned aggregation, yielding a spatially smooth and high-fidelity output.

\paragraph{Choice of spatial neural primitive}
We chose to use point-based neural primitives as the intermediate representation for 3D embedding, where the latent codes are anchored at the scene points. This point-anchor format is naturally suited for the multi-scene encoding via transformers and cross-scene generalization.
Several other neural graphics primitives have been proposed for compact scene representation and we acknowledge that there can be potentially better choices. Our preliminary experiments indicate that in single-scene optimization settings, point-anchored latent code displacements may not yet match the convergence efficiency of multi-resolution hash encodings~\cite{muller2022instant}. As illustrated in the figure (right), we evaluate this by making our per-point latent codes trainable and utilizing an MLP decoder to regress a single scene. Note that the relative simplicity of the MLP decoders in this experiment may result in lower visual fidelity compared to our primary results; however, they serve as a proof-of-concept for potential architectural extensions.

\setlength{\intextsep}{0pt}    
\setlength{\columnsep}{4pt}    
\begin{wrapfigure}{r}{0.3\textwidth}
\centering
\setlength{\tabcolsep}{0.5pt}
\begin{tabular}{
  >{\centering\arraybackslash}m{0.15\textwidth}
  >{\centering\arraybackslash}m{0.15\textwidth}
}
\includegraphics[width=0.15\textwidth]{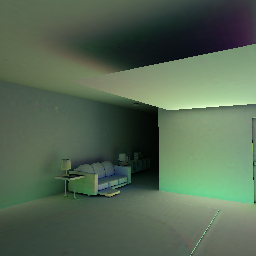} &
\includegraphics[width=0.15\textwidth]
{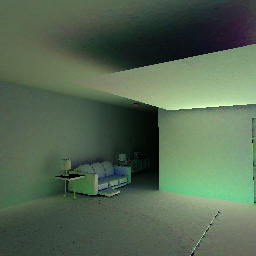}  \\
\footnotesize \shortstack{multi-res hashgrid} &
\footnotesize \shortstack{\textsf{trainable latent codes}\\\textsf{anchored at scene points}} \\
\end{tabular}
\end{wrapfigure}

Integrating our encoder with more expressive neural primitives remains a compelling direction for future work. For instance, adapting the encoder to produce latent features that index into a multi-resolution hash grid~\cite{muller2022instant} could potentially combine the benefits of transformer-based scene generalization with the high-resolution spatial encoding and accelerated inference characteristic of grid-based methods.

\paragraph{Scene point sampling strategy} We follow the work by \citet{hermosilla2018monte,li2019pointconv} to account for the importance sampling densities of points but find their impact on performance negligible. We observe that uniform sampling of 20k points per scene performs comparably to simple heuristics such as biasing toward smaller objects. In general, we find that the effect of sampling strategies diminishes once the scene is sufficiently covered. While performance is not highly sensitive to sampling strategy, the sampling density should still reflect geometric and texture frequencies to capture the signal accurately.

\paragraph{Number of query points for view-independent training}
To inform our design choices, we perform a quick verification experiment illustrated in \Cref{fig:ablation_query_points}. In this single-scene irradiance overfitting setup, we omit the encoder, make the point-based latent codes trainable, and use a simple MLP as the decoder. The experiment evaluates the effect of varying the number of query points. We find that using 2 million query points provides sufficient coverage for complex scenes, offering a good trade-off between performance and memory usage. For simpler scenes, this number can be significantly reduced; for instance, 200k points are sufficient for the Cornell Box.

\paragraph{Neighbor count $k$ and scene-point count $M$}
We further ablate on the number of input scene points and the neighborhood size $k$ used for the KNN search described in \S3.4.
We conduct these experiments on a small subset of the data, training on 90 scenes and testing on 10 scenes for easy analysis. 
As shown in Table~\ref{tab:hparams}, a small neighborhood ($k=8$) yields significantly lower performance, with accuracy saturating at $k \ge 32$. Given that larger $k$ values incur substantial computational overhead, we select $k=32$ as the optimal balance for our final model.

As expected, increasing the number of input scene points $M$ directly improves the performance, and our design choices are made with this in consideration (e.g. efficient light transport embedding \S3.3, local decoding \S3.4). We observe that beyond $M=20\text{k}$ points, PSNR improvements become marginal and visually imperceptible for our benchmark scenes. Consequently, we fix $M=20\text{k}$ as our default; however, for environments with higher geometric complexity or larger spatial extent, our architecture is designed to scale with increased point counts accordingly.




\begin{table}[t]
\centering
\caption{Ablation results for the number of neighbors (k) during decoding, and number of input scene points. Our choices are colored.}
\renewcommand{\arraystretch}{1.1}
\setlength{\tabcolsep}{4pt}
\begin{tabular}{>{\centering\arraybackslash}p{1.2cm}
                >{\centering\arraybackslash}p{1.6cm}
                !{\hspace{3pt}\vrule width 0.5pt\hspace{3pt}}
                >{\centering\arraybackslash}p{2.2cm}
                >{\centering\arraybackslash}p{1.6cm}}
\toprule
\textbf{k} & PSNR$\uparrow$ & \textbf{\# scene points} & PSNR$\uparrow$ \\
\midrule
8  & 26.96                    & 5k  & 28.51 \\
32 & \cellcolor{gray!20}34.57 & 10k & 30.94 \\
48 & 34.71                    & 20k & \cellcolor{gray!20}34.57 \\
   &                          & 40k & 36.25 \\
\bottomrule
\end{tabular}

\label{tab:hparams}
\end{table}
\paragraph{Embedding dimension $D$}
The hidden dimension $D$ defines the feature width across our entire architecture. Specifically, the Nearest-Neighbor Embedding projects aggregated tokens into a $D$-dimensional space; the PTV3 encoder operates within this space across all attention layers to produce the final light transport embeddings $\mathbf{F}_l$; and the Local Query Decoder performs cross-attention at the same dimensionality before final projection. Consequently, $D$ governs per-point representational capacity and maintains a synergistic relationship with $M$: while $M$ determines spatial coverage, $D$ dictates the richness of the features within that coverage. As shown in Table~\ref{tab:embed_dim}, doubling $D$ approximately doubles training VRAM usage at our current depth and patch size, with a commensurate increase in parameter count. At $D=256$, training memory requirements (43\,GB) exceed the capacity of standard 24\,GB GPUs, whereas $D=64$ provides insufficient per-token capacity for the scene complexity captured by $M$. We therefore select $D=128$ as our default, representing the largest dimension feasible on a single 24\,GB GPU. For future scaling, $D$ and $M$ should be scaled jointly to accommodate increasing scene complexity.

\begin{table}[h]
\vspace{10pt}
\centering
\caption{Effect of hidden dimension $D$ on model size and peak training VRAM (batch size 3, $M{=}20$k scene points, single GPU). Our default is highlighted.}

\small
\renewcommand{\arraystretch}{1.1}
\setlength{\tabcolsep}{8pt}
\begin{tabular}{ccc}
\toprule
$D$ & \textbf{\#Params} & \textbf{Train VRAM} \\
\midrule
64  & 10.5\,M  & 10.7\,GB \\
\cellcolor{gray!20}128 & \cellcolor{gray!20}42\,M  & \cellcolor{gray!20}21.3\,GB \\
256 & 167.7\,M & 43.1\,GB \\
\bottomrule
\end{tabular}
\label{tab:embed_dim}
\end{table}

\begin{figure}[!t]
\centering
\setlength{\tabcolsep}{1pt}
\renewcommand{\arraystretch}{1.0}
\begin{tabular}{
  >{\centering\arraybackslash}m{0.16\textwidth}
>{\centering\arraybackslash}m{0.16\textwidth}
  >{\centering\arraybackslash}m{0.16\textwidth}
}
\includegraphics[width=0.16\textwidth]{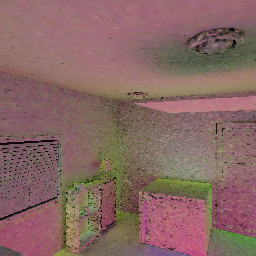} &
\includegraphics[width=0.16\textwidth]
{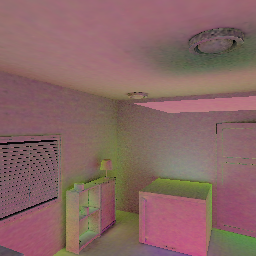} &
\includegraphics[width=0.16\textwidth]{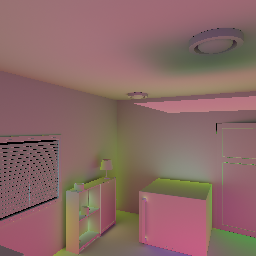} \\
\footnotesize \shortstack{262k query points} &
\footnotesize \shortstack{2M query points} &
\footnotesize \shortstack{Reference} \\
\end{tabular}
\caption{We analyze performance under varying numbers of sampled query points on the scene geometry in single-scene overfitting setting, where the model predicts irradiance at primary ray hits.}
\label{fig:ablation_query_points}
\end{figure}

\section{Extended baseline analysis}
As introduced in \S5.1 of the main paper, we compare our method against representatives from three distinct paradigms to cover the full spectrum of neural rendering. This broad evaluation complements recent work~\citet{zeng2025renderformer} by assessing performance across diverse architectural philosophies. Below, we elaborate on the experimental setup and discuss the results for each baseline in greater detail.

\paragraph{Deep Shading: screen-space limitations} The deep shading method \cite{nalbach2017deepshading} is confined to screen-space and has limited awareness of the actual 3D scene. As a result, it struggles to capture effects caused by out-of-sight geometries, exhibits view inconsistencies, and generalizes poorly to unseen scenes. 
To expose its limitations, we include the test scenes in the training set and show the resulting overfitted predictions in Figure 7.
For instance, accurately capturing the shadow cast by a light source onto the ceiling (indirect light bouncing from the floor) in \textsc{Dining Room 2} is not feasible without access to the 3D knowledge. By using direct lighting as an illumination cue, it was able to predict the area above the light source to be dark, while the region directly beneath it appears brightest (\textsc{Living Room}). We note that the screen-space CNN paradigm is also adopted by \citet{xin2022lightweight}; we use Deep Shading as a representative of this category since both share the same fundamental view-dependence and off-screen-blindness. Recent advances in video foundation models may offer improved spatial reasoning by leveraging the 2D+time dimension. However, with the increasing availability of large-scale 3D data, it may be more effective to bypass the indirect route of reconstructing 3D from 2D projections and instead embrace direct 3D representations—enabling full editability and consistent understanding, as demonstrated by our method.

\paragraph{\citet{hermosilla2019deep}: point convolutions on complex scenes} In general, we observe that transformers exhibit strong generalization ability and are surprisingly resistant to overfitting, even on small datasets. In contrast, \citet{hermosilla2019deep} is built upon point convolutions, which struggle to learn effectively from our complex dataset. Although we modified their model to better match our scene complexity, it still fails to capture the intricate nature of global illumination and suffers from noticeable artifacts. We believe transformers offer an advantage due to their ability to model long-range dependencies and dynamically aggregate features across spatially distant regions, which is critical for accurately representing light transport.

\paragraph{Note on path tracing as a baseline} While path tracing often produces noisy outputs, these are typically refined using post-processing denoisers in modern pipelines. We do not claim our method replaces the full path tracing plus denoising setup, but instead emphasize the potential of predicting global illumination directly from 3D scene configurations. Unlike traditional path tracing, where shader execution diverges across pixels due to bounce paths, occlusion, and material interactions, our method performs a uniform forward pass per pixel. This regularity enables efficient parallel execution on modern GPUs and offers predictable performance across varying scene complexity. It also replaces the irregular, memory-bound nature of path tracing with more streamlined and predictable memory access. As power efficiency becomes increasingly critical in modern architectures, transformer-based inference presents a promising direction for scalable light transport.

\paragraph{Discussion on RenderFormer} 
To further contextualize our contributions, we provide a detailed comparison with RenderFormer, which represents the state-of-the-art in transformer-based global illumination for object-centric tasks. Beyond the scalability differences summarized in Table 2 of the main paper, we expand on the functional capabilities and operational constraints of both methods in Table~\ref{tab:capabilities}. The two methods diverge fundamentally in their supervision signals and scene representations.


\begin{table}[t]
\vspace{5pt}
\centering
\caption{Capability scope. We demonstrate the capabilities of our method as shown in our results, comparing them with RenderFormer~\cite{zeng2025renderformer} across four key axes: scene complexity, rendering resolution, texture fidelity, and camera freedom.}
\label{tab:capabilities}
\small
\setlength{\tabcolsep}{2pt}
\begin{tabular}{l p{6.5cm}}
\toprule
\textbf{Feature} & \textbf{Ours vs. RenderFormer} \\
\midrule
{Scene Complexity} & 
 RenderFormer is limited to $\approx$\textbf{4k} triangles due to $\mathcal{O}(M^2)$ attention costs and the coupling with image resolutions, often requiring decimation that causes geometry loss. In contrast, our method scales via linear attention; our evaluation scenes average \textbf{3.2M} primitives (see Table~\ref{tab:scene_stats}). \\
\midrule
{Resolution} & 
RenderFormer is tied to its fixed training resolution (\textbf{512$\times$512}) because queries are coupled by the vision transformer and the resource conflict with geometries. Our method is independent from resolution; Figure 1 demonstrates \textbf{2688$\times$1152} resolution with consistent quality.\\
\midrule
{Textures} & 
RenderFormer assumes constant reflectance per triangle or limited $32{\times}32$ textures. Our method supports high-fidelity textures, averaging 40 textures per scene at \textbf{1024$^2$} resolution (see Table~\ref{tab:scene_stats}). \\
\midrule
{Camera Space} & 
 We support immersive ``walk-throughs'' with cameras placed anywhere, while RenderFormer restricts the camera to outside the cornell bounding box. \\
 \midrule
{Material effects} & 
 RenderFormer achieves impressive results on high-frequency specular materials by directly predicting outgoing radiance. In contrast, we prioritize view consistency and resolution independence; thus, we predict incident radiance using 3D supervision, which ensures stability across arbitrary camera paths at the cost of some specular sharpness. \\
\bottomrule
\end{tabular}
\end{table}

\begin{table}[h]
\vspace{10pt}
\centering
\caption{{Example dataset statistics.} We show the statistics of our complex indoor scenes exhibiting high geometric density and high-resolution textures, far exceeding the $\approx$4k primitives limit of \citet{zeng2025renderformer}.}
\label{tab:scene_stats}
\small
\begin{tabular}{l c c c}
\toprule
\textbf{Scene Figure} & \textbf{Triangle Count}  & \textbf{Texture Resolution}\\
\midrule
\textit{Teaser Figure 1} & 3,427,263 & 46 $\times 1024^2$  \\
\textit{Bath Room Figure 7} & 825,645  & 29 $\times 1024^2$ \\
\textit{Hall Way Figure 7} & 2,459,417 & 41 $\times 1024^2$  \\
\textit{Dining Room 1 Figure 7} & 2,765,279 & 35 $\times 1024^2$  \\
\textit{Dining Room 2 Figure 7} & 4,622,312 & 44 $\times 1024^2$  \\
\textit{Living Room Figure 7} & 2,711,230 & 51 $\times 1024^2$ \\
\textit{Study Room Figure 7} & 6,044,118 & 29 $\times 1024^2$ \\
\bottomrule
\end{tabular}
\end{table}


For the qualitative comparison in Figure 10 of the main paper, we took extensive measures to adapt the standard Cornell Box to RenderFormer's specific input assumptions before invoking the released checkpoint. Geometry was tessellated to match their templates. PBRT's diffuse walls map directly to RenderFormer's diffuse term; the PBRT Mirror conductor ($\eta + ik$ with roughness 0.1) is mapped to specular $\approx 0.9$ (the Fresnel $F_0$ at normal incidence) with roughness passed through, plus a small diffuse so that diffuse + specular falls within RenderFormer's training range of $[0.9, 1.0]$. The original area light is replaced by a single emissive triangle at the same centroid, sized to match RenderFormer's training-distribution emitters (similar to point lights) and with radiance set to preserve the original total flux. The flux-correct radiance falls below RenderFormer's training range of $L \in [2500, 5000]\,\mathrm{W/units^2}$, and RenderFormer does not appear to extrapolate linearly outside that range: below 2500 it under-produces light, contributing to the dimmer appearance of its panel. Improving training efficiency by enforcing such physical invariances (e.g., linear radiance scaling) remains an area for future work.

Our model was fine-tuned for one day on a small, low-poly subset to verify its capability in this secondary low-poly regime. While the limited training diversity in this subset results in minor shadow blotchiness, these artifacts are expected to resolve with more comprehensive training on a larger dataset.


\section{Additional results and applications}

\subsection{Inherent view consistency}
Our model is inherently view-independent (\S4 of main paper), as it is trained on query points sampled directly from the 3D scene rather than camera-based observations. Please refer to our supplementary video. 

\subsection{Scene editing}
\begin{figure}[!t]
    \centering
    \includegraphics[width=1.0\linewidth]{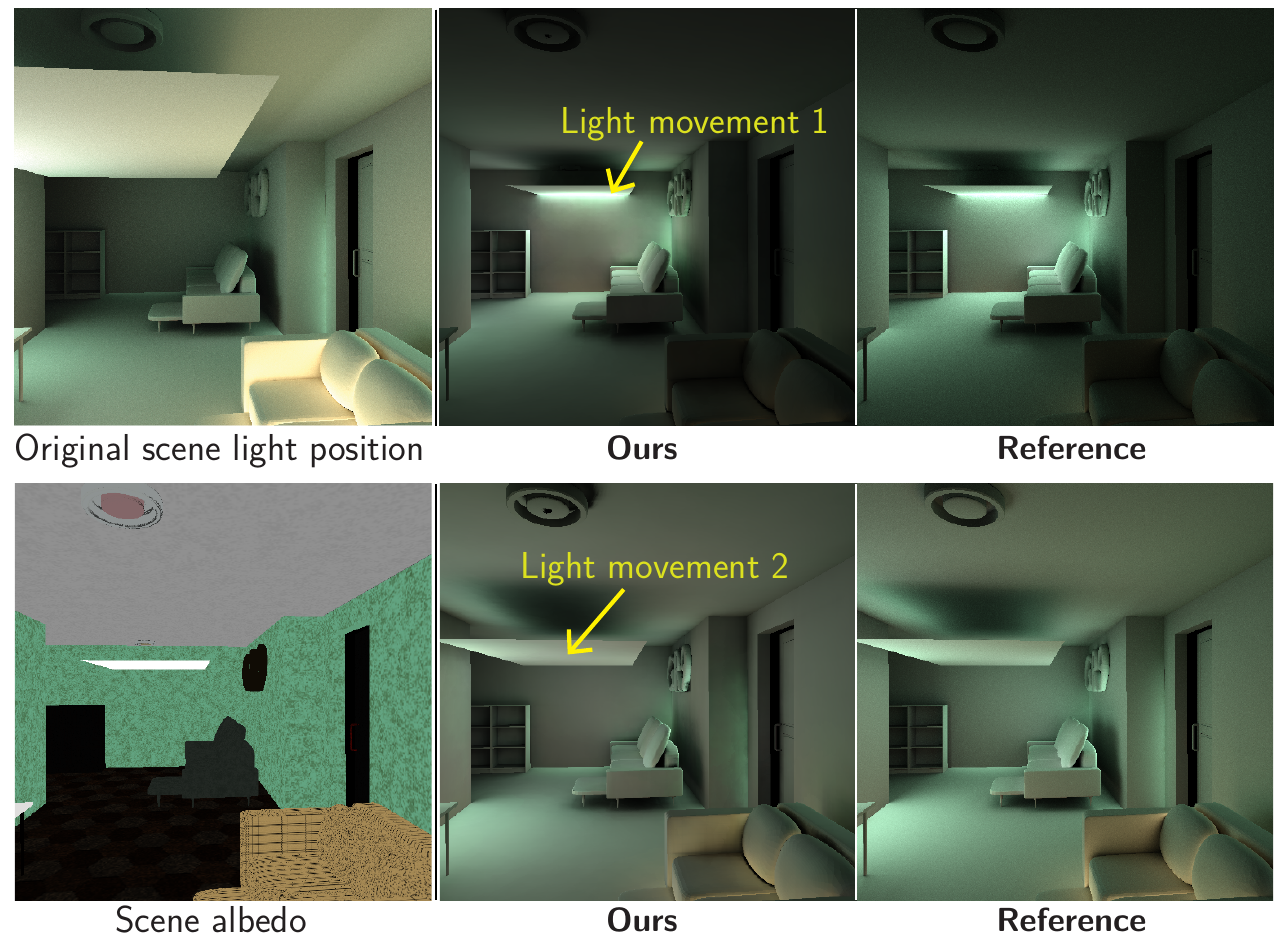}
    \caption{As the area light moves away or toward the camera, both the shadow and indirect illumination respond accordingly. The model generalizes reasonably well to novel light placements within the same scene. We expect that increasing the diversity of light positions during training would further enhance the quality and robustness of the results. Again, we show the corresponding albedo to provide readers context for the color of the indirect lighting, while we avoid showing albedo-modulated images to maintain a clear focus on the quality of the predicted irradiance without distraction. }
    \label{fig:moving_lights}
\end{figure}

\paragraph{Light source movement}
We assess the model's robustness to changing light positions, as illustrated in \cref{fig:moving_lights}. In this living room example, moving the area light toward the camera leads to corresponding changes in shadows and indirect illumination. The shadow shifts relative to the light source due to occlusion of indirect light reflected from the floor, and the two sofa areas become brighter or darker as a result. Although each scene was trained with a fixed light configuration, the model not only generalizes to new floor plans and geometries, but also responds robustly to novel light placements. This behavior reflects the strength of the learned 3D light transport embedding in capturing global illumination dynamics. We expect that training on a wider range of lighting conditions would further enhance both accuracy and robustness.

\begin{figure}[!t]
    \centering
    \includegraphics[width=1.0\linewidth]{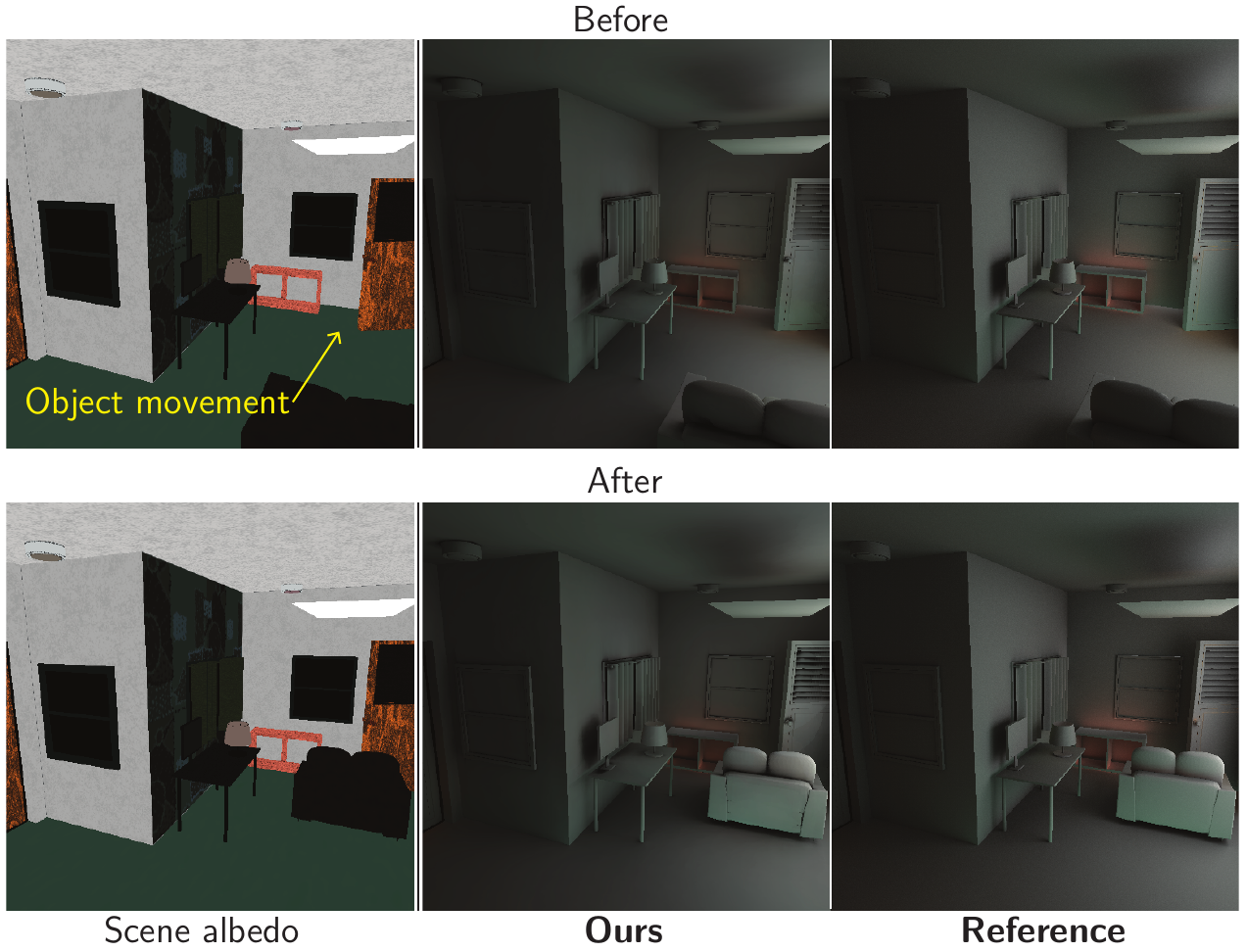}
    \caption{Irradiance prediction under non-emissive object movement. Unlike previous works, no per-scene training is performed. As the sofa moves toward the light source, it receives stronger indirect illumination, while nearby regions—such as the cabinet near the wall—become darker due to increased occlusion. Again, we show albedo separately to maintain a clear focus on the quality of the predicted irradiance without distraction.}
    \label{fig:moving_objs}
\end{figure}

\paragraph{Object movement} 
We evaluate the model’s robustness to non-emissive object movement, as shown in \cref{fig:moving_objs}. In this example, as the sofa moves closer to the light source, it becomes more brightly lit, while the small cabinet near the wall appears darker due to increased occlusion. As expected, the shadow consistently follows the moving object. 

\begin{figure}[!t]
    \centering
    \includegraphics[width=1.0\linewidth]{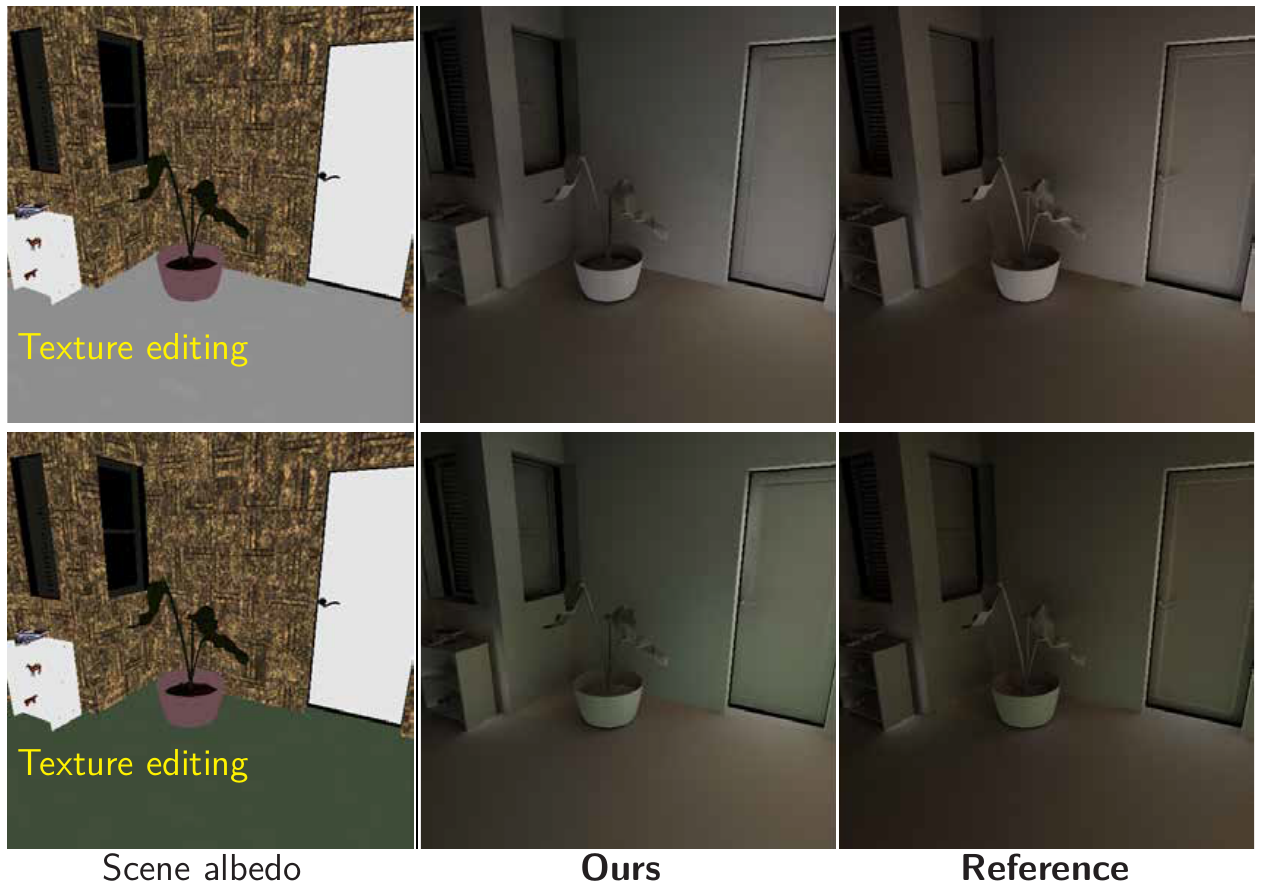}
    \caption{Irradiance prediction under material editing. When we change the floor’s albedo from a pale (whitish) tone to a greenish hue, the indirect lighting cast onto the walls adjusts correctly. No per-scene retraining is needed.}
    \label{fig:material_edit}
\end{figure}

\paragraph{Material editing}
We evaluate the model's response to material editing. In \cref{fig:material_edit}, changing the floor’s albedo from pale white to deep green causes the predicted irradiance field to update accordingly, modifying the indirect lighting on nearby walls. This result is achieved without any per-scene retraining. Expanding the dataset with a wider range of textures would likely improve robustness in more diverse editing scenarios.

\subsection{Architectural versatility and directional bases}
\label{supp:sh}
While our primary focus is on scalable diffuse transport, the underlying 3D embedding is designed to be task-agnostic. In this section, we expand on the architectural extension introduced in \S5.3 of the main paper, focusing on the choice of directional representation for incident radiance.

\paragraph{Learned histogram vs.\ spherical harmonics}
We evaluate our neural directional basis against Spherical Harmonics (SH), the standard parametric basis in neural rendering. While SH offers a compact encoding, it faces a fundamental trade-off between smoothness and frequency resolution. 
Low-degree SH basis functions tend to oversmooth sharp structures, whereas higher degrees $(l_{\max} \ge 4)$ significantly increase the feature dimensionality—$(l_{\max}+1)^2$ coefficients per color channel—and often introduce Gibbs-phenomenon ringing artifacts near high-frequency boundaries.

By conditioning on scene geometry and material context, our learned basis achieves significantly higher fidelity for complex incident-radiance distributions.
Qualitative comparisons and error vs.\ degree \(l_{\max}\) are reported in \cref{fig:neural_basis}, where our method preserves sharp detail without ringing and attains lower reconstruction error.

\paragraph{Future extension: parametric neural bases}
While the current histogram implementation is more expressive for high-frequency structures, it remains discrete. A natural progression for our framework is to utilize the neural conditioning to predict parameters for continuous bases, such as Spherical Gaussians (SG). This would allow for a more compact representation while maintaining the resolution-independence and view-consistency of our 3D embedding.

\begin{figure}[!t]
    \centering
    \includegraphics[width=1.0\linewidth]{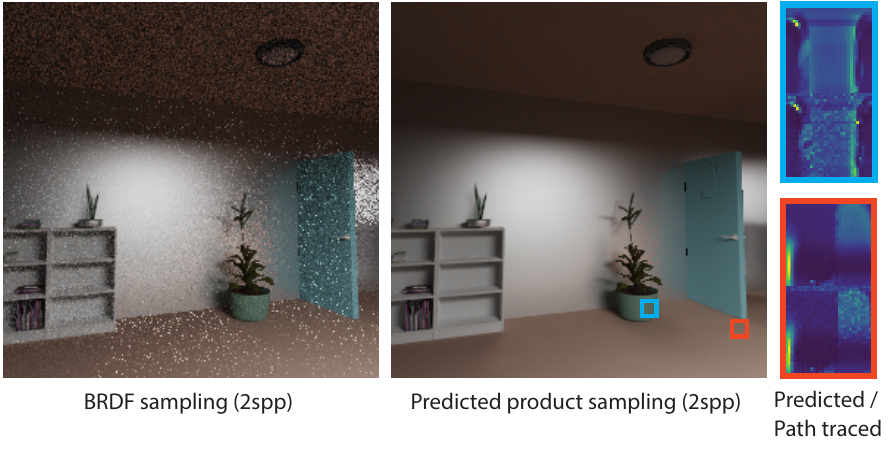}
    \caption{Path tracing results using different sampling strategies. Left: BRDF sampling with 2 samples per pixel (spp); middle: importance sampling via CDF inversion based on the normalized histogram of the product of our predicted incoming radiance field and brdf values; right: the predicted (top) and path traced (bottom) sampling distributions for the incident radiance integrand at two random points. The reference can be seen in Figure 9 of the main paper. Our model-driven sampling significantly reduces noise in low-sample regimes. As more samples are accumulated, our predicted PDF can serve as an effective initializer for modern path guiding methods, mitigating the cold-start problem inherent in per-scene optimization approaches. Note that the cosine term is included in both of the importance sampling strategies.}
    \label{fig:path_guiding}
\end{figure}

\begin{figure*}[!t]
    \centering
    \includegraphics[width=1.0\linewidth]{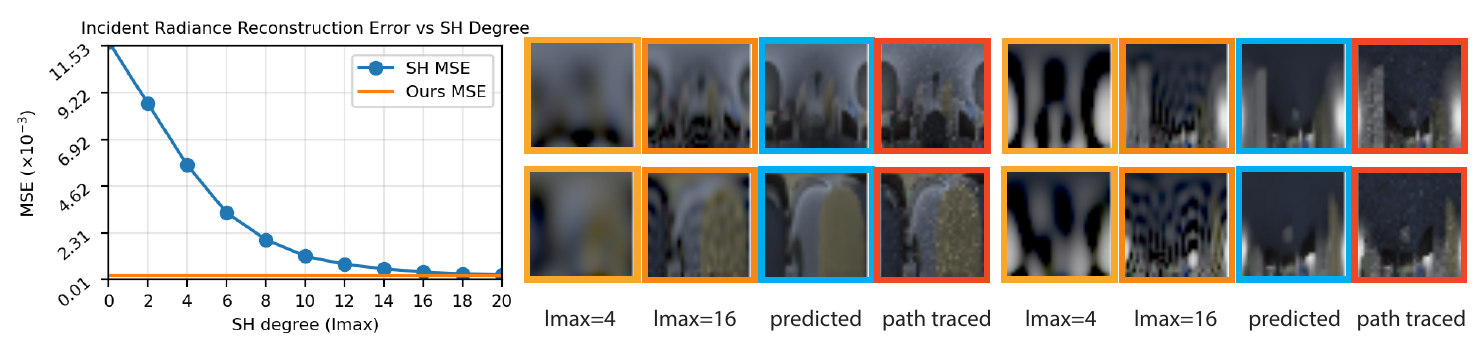}
    \caption{ To assess the effectiveness of our scene encoder and neural basis, we compare our incident-radiance predictions against spherical-harmonics (SH) reconstructions at four randomly sampled surface points. Our method preserves high-frequency glossy reflections. In contrast, low-degree SH yields overly blurred results (first column), while high-degree SH (second column) introduces ringing (Gibbs-like) artifacts. The left plot reports the reconstruction error (MSE) as the SH degree $l_{max}$ increases.}
    \hspace{4pt}
    \label{fig:neural_basis}
\end{figure*}

\subsection{Jump-starting path guiding}
\label{sec:importance_sampling}
While the two previous applications focus on directly predicting global illumination, we provide an initial exploration into how our learned embeddings might support traditional unbiased rendering by aiding importance sampling.

Building on the predicted directional radiance field from \S5.3 of the main paper, we explore its use in path guiding initialization. As shown in \Cref{fig:path_guiding}, we compare standard BRDF sampling against importance sampling derived from the product of our predicted radiance field and the BRDF term via CDF inversion. To maintain the unbiased nature of the estimator, the predicted guiding PDF is designed with full support (i.e., non-zero probability for every non-zero bin), ensuring that mathematical correctness is preserved even when using an approximate distribution. At very low sample counts (e.g., 2 samples per pixel), modern path guiding methods (\cite{pathguidingcourse}, \cite{pathguidingcourse2025}) typically lack sufficient samples to construct a usable distribution. Our learned radiance field serves as a ``cold-start'' initializer, effectively ``jump-starting'' the guiding process by providing an informed distribution before scene-specific data is gathered. Our model is not intended as a standalone path-guiding solution, but rather as a prior for established methods that utilize defensive BRDF sampling. This approach draws parallels with meta-learning, where knowledge accumulated across scenes enables fast adaptation to new instances, facilitating efficient light transport simulation without per-scene retraining. 

We emphasize that this application is intended as a proof-of-concept; the network currently predicts angularly smooth signals via $32 \times 32$ histogram supervision, where each bin represents an average over its respective solid-angle subdomain. Consequently, the result is a low-resolution approximation of the true radiance distribution with an inherent frequency limit. Furthermore, we acknowledge that invoking the full MLP decoder to query directional quantities for each individual sample is currently computationally redundant. This per-query overhead could be bypassed in future work by exploring more efficient pathways to amortize the decoding cost, such as utilizing a CNN-based head to predict a direct bitmap of the radiance distribution or predicting basis coefficients in a single forward pass. 

\end{document}